\documentclass[preprint,showpacs,amssymb,amsmath,nofootinbib,longbibliography]{revtex4-1}

\usepackage{color}
\usepackage[dvipdfmx]{graphicx}
\usepackage{dcolumn}
\usepackage{bm}
\usepackage{ulem}

\begin{document}

\title{
All-mode Renormalization for Tensor Network \\ with Stochastic Noise 
\vspace{10mm} }

\newcommand{\Rikenbnl}{
RIKEN BNL Research Center, Brookhaven National Laboratory, Upton New York 11973, USA
}
\newcommand{\NaraWU}{
Department of Physics, Nara Women's University, Nara 630-8506, Japan
}
\newcommand{\UCONN}{
Physics Department, University of Connecticut, Storrs, CT 06269-3046, USA
}
\newcommand{\Kanazawa}{
Institute for Theoretical Physics, Kanazawa University,
Kanazawa 920-1192, Japan
}

\author{Erika Arai}
\affiliation{\NaraWU}

\author{Hiroshi Ohki}
\affiliation{\NaraWU}
\affiliation{\Rikenbnl}

\author{Shinji Takeda}
\affiliation{\Kanazawa} 

\author{Masaaki Tomii}
\affiliation{\UCONN}

\begin{abstract}

In usual (nonstochastic) tensor network calculations, the truncated singular value decomposition is often used for approximating a tensor,
and it causes systematic errors.
By introducing stochastic noise in the approximation, however, 
one can avoid such systematic errors at the expense of statistical errors which can be straightforwardly controlled.
Therefore in principle, exact results can be obtained even at finite bond dimension up to the statistical errors.
A previous study of the unbiased method implemented in tensor renormalization group algorithm, however, showed that
the statistical errors for physical quantity are not negligible, and
furthermore the computational cost is linearly proportional to a system volume.
In this paper, we introduce a new way of stochastic noise such that the statistical error is suppressed, and
moreover, in order to reduce the computational cost we propose common noise method whose cost is proportional to the logarithm of volume.
We find that the method provides better accuracy for the free energy compared with 
the truncated singular value decomposition when applying to tensor renormalization group for Ising model on square lattice.
Although the common noise method introduces systematic error originated from a correlation of noises,
we show that the error can be described by a simple functional form in terms of the number of noises, thus
the error can be straightforwardly controlled in an actual analysis.
We also apply the method to 
the graph independent local truncation
algorithm and show that the accuracy is further improved.

\end{abstract}

\preprint{KANAZAWA-22-06}

\maketitle

\section{\label{sec:intro}Introduction}
\noindent
A numerical renormalization group (RG) technique based on the tensor networks 
becomes a popular and powerful tool to study the quantum lattice field theory and 
the many-body systems in condensed matter physics.  
A typical and simple algorithm for a tensor contraction is the tensor renormalization group (TRG) approach~\cite{Levin:2006jai},  
where the tensor contraction is performed using the low-rank approximation via the singular value decomposition (SVD).
The lattice topology is unchanged after a tensor reconstruction step, thus
this coarse graining procedure can be repeated and a large volume simulation can be easily realized. 
Another striking feature of the TRG is that 
there is no sign problem 
and hence TRG-related methods can be applied 
to systems such as a complex action with $\theta$ term, 
the real-time dynamics that are not easily accessible by the Monte Carlo 
methods~\cite{Shimizu:2014uva, 
Hong:2022iwh, 
Bazavov:2019qih, 
Kadoh:2018hqq, 
Kadoh:2018tis, 
Yoshimura:2017jpk, 
Kawauchi:2016xng, 
Kuramashi:2019cgs, 
jahromi2018infinite, 
Sakai:2017jwp, 
Shimizu:2014fsa, 
Shimizu:2017onf, 
Takeda:2014vwa}.\footnote{For recent lattice studies of tensor network 
with applied quantum computing and for a system with higher dimensions, 
see reviews \cite{Banuls:2019bmf, Kadoh:2022loj} and references therein.}

A key ingredient of the TRG is using the truncated SVD when decomposing a tensor,
and this plays an important role to realize an efficient and sustainable coarse-graining algorithm.
The truncation of the lower modes, however, causes a systematic error, and 
in general it is difficult to predict a scaling property of the truncation error especially when repeating several coarse-graining steps.
Such a truncation technique is commonly used in the improved coarse-graining algorithms
such as Tensor Network Renormalization (TNR)~\cite{evenbly2015tensor}, loop-TNR~\cite{yang2017loop}, Gilt~\cite{Hauru:2017jbf} 
and also
in the variety of efficient cost reduction algorithms~\cite{Xie:2009zzd, xie2012coarse, zhao2016tensor, 
wang2014phase, harada2018entanglement, morita2018tensor, evenbly2017algorithms, Nakamura:2018enp, Adachi:2019paf, Kadoh:2019kqk, Lan:2019stu, Oba:2019csk, Morita:2020com, Adachi:2020upk, Banuls:2019hzc}. 
In fact, for high precision calculations a careful treatment of the systematic error estimate is needed, 
so that it is important to pursue a possibility of alternative algorithms 
that can improve the error evaluation method as well as the numerical accuracy itself.  
One of possibilities to remove the systematic error is to use 
the Monte Carlo tensor network (MCTN)~\cite{ferris2015unbiased, 2017arXiv171003757H},
in which a Monte Carlo sampling of the singular modes is employed 
and there is only a statistical error. 
For different Monte Carlo approaches in variational methods in a tensor network representation
see \cite{sikora2015variational, sandvik2007variational, wang2011monte, ferris2012perfect, ferris2012variational, Iblisdir, schwarz2017projector, zhao2017variational, Vieijra:2021bis, qin2020combination}.

In this work, following the strategy of the MCTN 
we propose a new stochastic method for tensor decomposition and apply it to the TRG. 
A basic idea of using the stochastic method 
is as follows.
In the usual truncated SVD, the lower modes are just discarded, and
such a low rank approximation is known to be the best one from a local point of view but not necessarily for global quantity like a partition function.
The contribution of the lower modes, however, can be incorporated into decomposed tensors by combining them with stochastic noises.
By using the tensors which are compact but contain all singular modes (all-mode renormalization), the coarse graining can be done as in the case of the TRG.
Note that in the stochastic method the truncation error can be replaced with an statistical error due to the noise, 
and a stochastic determination of the partition function and any related physical quantities is possible.  
We examine two types of spatial distribution of noise: position-dependent and -independent 
ways (for the latter case we refer it as the common noise method).
As for the position-dependent noise method it is applicable to the system that has no translation invariance 
thus the computational complexity scales as its volume. 
In this case there is no systematic error and it provides an unbiased result,
which will be directly confirmed by a numerical calculation. 
As for the common noise method,  
the order of the computational cost remains the same as the TRG,  
while as will be discussed there is a noise cross contamination effect, 
which turns out to be the only source of the residual systematic error. 
Nevertheless we find a significant improvement of the accuracy compared to the TRG. 
Since this residual systematic error has a simple scaling property due to a nature of the random noise, 
we provide a systematic 
error evaluation method, 
which is independent of 
model dynamics and tensor network algorithms.
Thus our new stochastic noise method combining with the deterministic algorithm
can actually improve the error estimation 
 as well as the numerical accuracy. 
Moreover, our method is so simple that can easily be applied to a complicated system and combined with improved tensor network algorithms.

Our method shares the same idea as the MCTN, 
but there are some practically important differences.
The MCTN uses a subset of the singular modes for a tensor decomposition that are randomly chosen 
with an appropriate probability distribution (see \cite{ferris2015unbiased} for details),
while in our case all the singular modes are manifestly included thanks to the random noise vectors.
Another important difference is that we propose the common noise method that is not considered in \cite{ferris2015unbiased, 2017arXiv171003757H},
where the position-dependent method in our language is only proposed and examined.
As mentioned before, the common noise method has the desired properties: it provides better accuracy, the residual systematic error is under control,
and the computational cost is the same order as the TRG.
Therefore we consider that the common noise method is practically useful for future applications.

The paper is organized as follows. In Sec.~\ref{sec:noise} we review the TRG algorithm and introduce 
a tensor decomposition method using noise vectors. 
In Sec.~\ref{sec:hybrid} after introducing an ensemble method, 
we define the position-dependent and the common noise method.
We test the noise methods for the 2D Ising model in Sec.~\ref{sec:test}.
We discuss a possible application to other TRG algorithms and show some numerical results when applying to the Gilt in Sec~\ref{sec:sGilt-TNR}.
Section \ref{sec:conclusion} is devoted to the conclusion. 
In the Appendix, the results for the specific heat are shown as an example of related thermodynamical quantities.
Our preliminary result has been published in \cite{Ohki:2021ukk}.

\section{\label{sec:noise}Tensor decomposition with random noise vector}

\begin{figure}[tbph]
\begin{center}
\includegraphics[clip,width=0.45\textwidth]{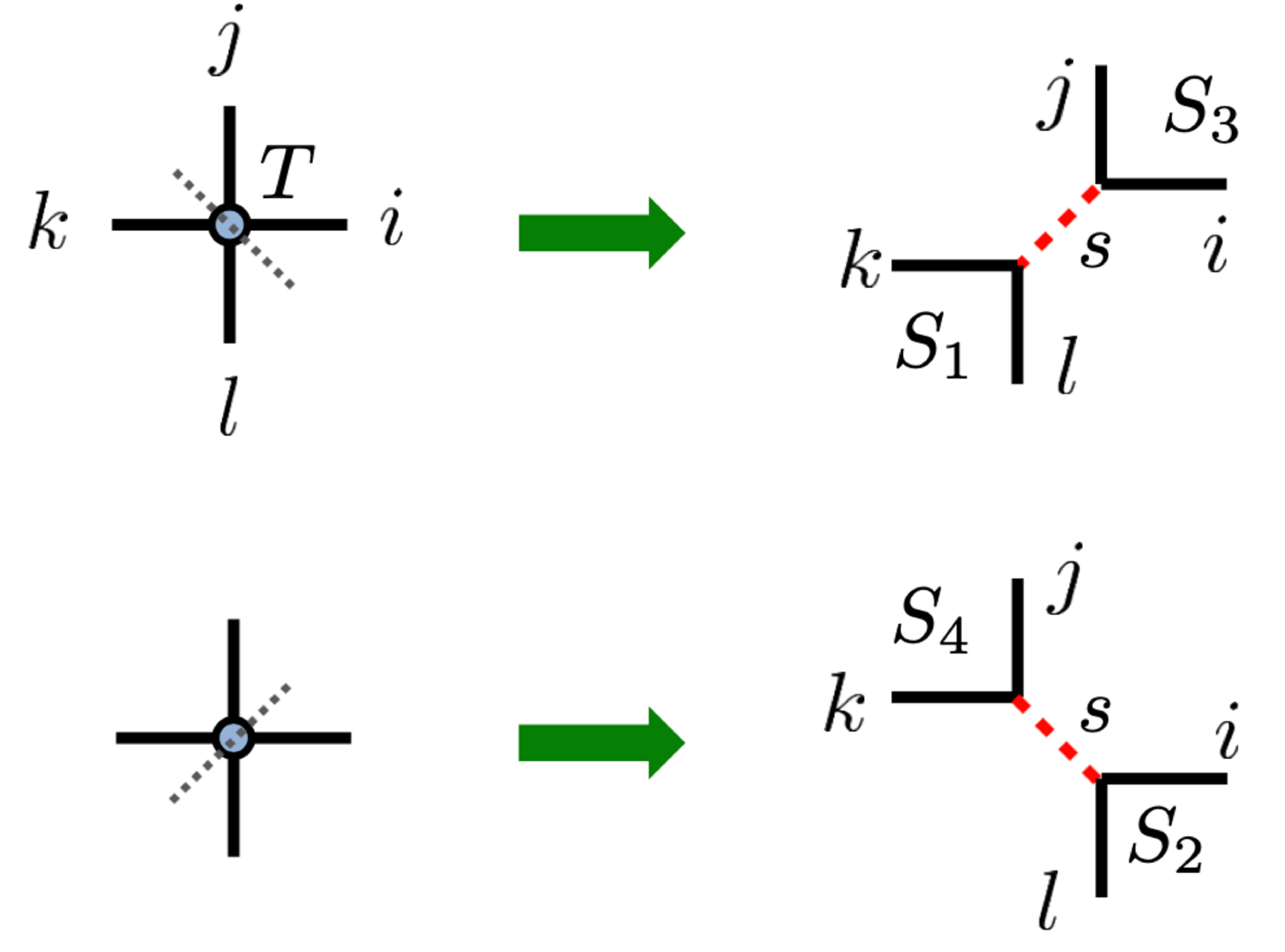} 
\end{center}
\caption{\label{fig:TRG_decomposition} 
Decomposition of a tensor $T_{ijkl}$ to two third-order tensors $S_{3as}S_{1sb}$ (upper panel)
and $S_{4cs}S_{2sd}$ (lower panel).}
\end{figure}

\begin{figure}[tbph]
\begin{center}
\includegraphics[clip,width=0.5\textwidth]{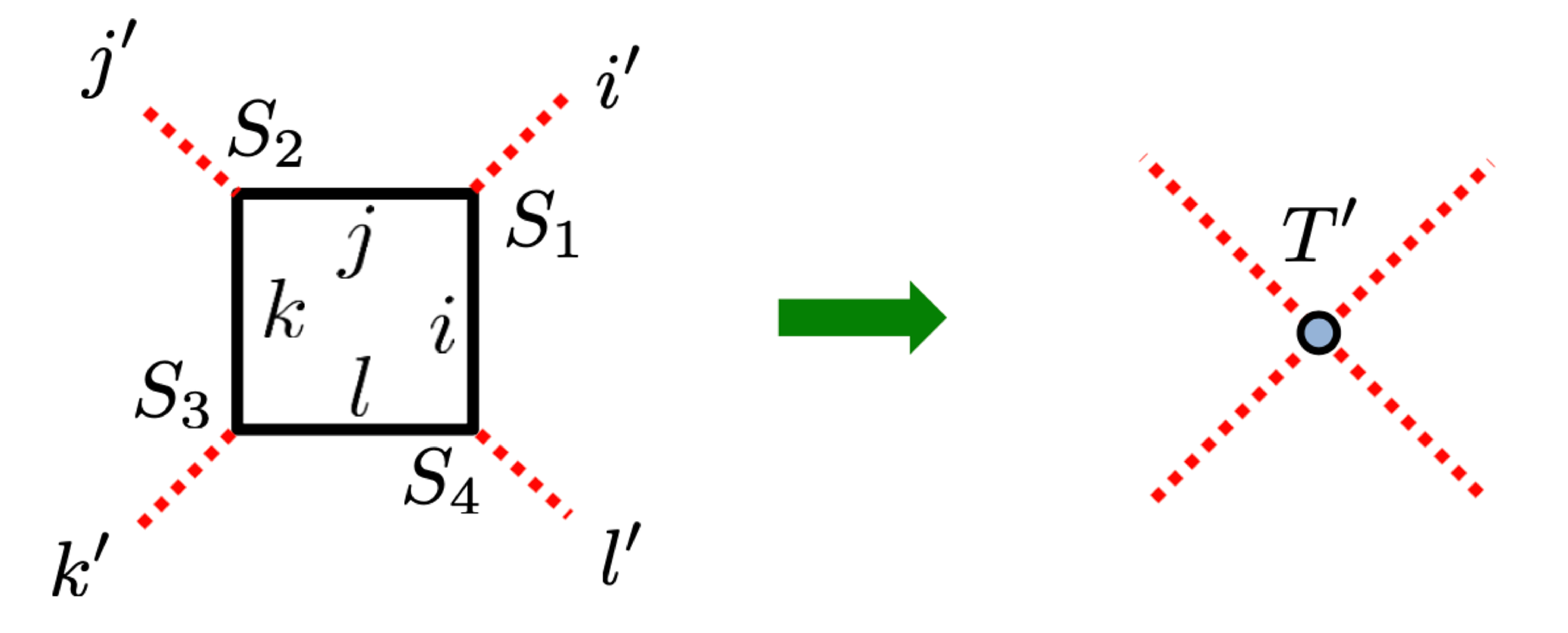} 
\end{center}
\caption{\label{fig:TRG_contraction} 
Contraction to create a coarse grained tensor $T^\prime_{i^\prime j^\prime k^\prime l^\prime}$ from the third-order tensors $S_{1,2,3,4}$.}
\end{figure}

We first review the TRG method for the 2D Ising model on a square lattice. 
The partition function $Z$ for the model is given as a trace over tensor indices in a tensor network representation,  
\begin{align}
Z\equiv \sum_{\{\sigma\}} e^{-\beta H[\sigma]} = \sum_{ijk\cdots}T_{ijkl}T_{mnip}\cdots = {\rm Tr}\left[\otimes T \right], 
\end{align}
where $T_{ijkl}$ is the initial tensor. 
We express the tensor in a matrix representation, $T_{ijkl} = M_{ij;kl} \equiv M_{ab}$, and $T_{ijkl} = \tilde{M}_{jk;li} \equiv \tilde{M}_{cd}$ with different combination of the indices. 
Using the SVD, we represent a matrix $M$ with a rank $R$ as follows:
\begin{align} \label{eq:fullSVD}
M_{ab} = \sum_{s=1}^R \sqrt{\Lambda_s} u_{as} \sqrt{\Lambda_s} v_{sb}, 
\end{align}
where $\Lambda_s$ are singular values ($\Lambda_1 \geq \Lambda_2 \geq \Lambda_3 \geq \cdots$).
By truncating the lower modes and keeping the largest $D_{\rm svd} (\leq R)$ mode,
the fourth-order tensor is approximated by a product of two third-order tensors 
\begin{align}
\label{eq:SVD}
T_{ijkl} = M_{ab} \simeq \sum_{s=1}^{D_{\rm svd}} \sqrt{\Lambda_s} u_{as} \sqrt{\Lambda_s} v_{sb} = 
\sum_{s=1}^{D_{\rm svd}} S_{3as}S_{1sb},
\end{align}
where $S_{3as}=\sqrt{\Lambda_s} u_{as}$ and $S_{1sb}=\sqrt{\Lambda_s} v_{sb}$ 
are the third-order tensors. 
Similarly $S_{2,4}$ are also defined to approximate the other type of a tensor decomposition, 
$\displaystyle \tilde{M}_{cd} \simeq \sum_{s=1}^{D_{\rm svd}} S_{4cs} S_{2sd}$.
See Fig.~\ref{fig:TRG_decomposition} for pictorial expression of the decompositions.
We then obtain a coarse grained tensor by contracting 
all old indices ($i,j,k,l$) of the four third-order tensors of $S_{1,2,3,4}$ (see Fig.~\ref{fig:TRG_contraction})
\begin{align}
T^{\prime}_{i'j'k'l'} = \sum_{ijkl} S_{1i';ji}S_{2j';kj} S_{3lk;k'} S_{4il;l'}
\equiv {\rm Tr}[ S_{1i'}S_{2j'} S_{3k'} S_{4l'}],
\end{align}
where ${\rm Tr}$ stands for contracting all old indices.
By repeating
the coarse-graining $n$-times, we obtain a renormalized tensor $T^{(n)}_{i'j'k'l'}$ as
\begin{align}
T^{(n)}_{i'j'k'l'} 
= {\rm Tr}[ S^{(n-1)}_{1i'}S^{(n-1)}_{2j'} S^{(n-1)}_{3k'} S^{(n-1)}_{4l'}], \ \ (n >0)
\end{align}
where $T^{(0)}=T$ and $S^{(0)}_{1,2,3,4}=S_{1,2,3,4}$.
The bond dimension of the renormalized tensor is $D_{\rm svd}$ and 
the dominant computational cost for contractions scales as $\mathcal{O}(D_{\rm svd}^6)$. 


Next let us explain our noise method. 
In the tensor decomposition step in the TRG, the truncated SVD is used,
and it causes systematic errors.
In order to reduce the systematic uncertainty with a limited number of bond dimensions, 
we introduce $D$-dimensional noise vectors $\eta_r=(\eta_{1r}, \eta_{2r}, \cdots, \eta_{Dr} )^T$ for $r=1,\cdots, N_r$, 
where each component $\eta_{ir}$ is an element of $Z_N$. 
We then define a $D \times N_r$ matrix 
$\bm{\eta} = (\eta_1, \cdots, \eta_{N_r})$, which satisfies the completeness condition 
\begin{align}
\label{eq:noise}
\frac{1}{N_r} {\bm\eta} \cdot {\bm\eta}^\dagger = \frac{1}{N_r} \sum_{r=1}^{N_r} \eta_r \cdot \eta_r^\dagger =
{\bm 1}_{D\times D} + \mathcal{O}(1/\sqrt{N_r}),
\end{align}
where we note that for $Z_N$ noise the diagonal part is exactly unity and the statistical fluctuation
appears only in the off-diagonal part.
We use the random noise vector\footnote{
Although $\bm{\eta}$ is a matrix, it is convention to call it noise vector.
} in the decomposition of a matrix using SVD.
Substituting ${\bm 1}_{D\times D} \simeq \frac{1}{N_r}{\bm \eta}\cdot{\bm \eta}^\dagger$ into the lower mode subspace in Eq.~\eqref{eq:fullSVD} 
with $D=R-D_{\rm svd}$, the matrix $M$ is given by 
\begin{align} \notag
M_{ab} 
=
& \sum_{s=1}^{D_{\rm svd}} (\sqrt{\Lambda_s} u_{as} \sqrt{\Lambda_s} v_{sb}) 
+ \sum_{s,t=D_{\rm svd}+1}^R 
\sqrt{\Lambda_s} u_{as} \left(\frac{1}{N_r}{\bm \eta}\cdot{\bm \eta}^\dagger\right)_{s-D_{\rm svd},t-D_{\rm svd}} \sqrt{\Lambda_t} v_{tb}
\\ 
&+ \mathcal{O} \left(\sqrt{\frac{\Lambda_{D_{\rm svd}+1} \Lambda_{D_{\rm svd}+2}}{N_r}} \right),
\end{align}
where the first term represents the contributions from the largest $D_{\rm svd}$ singular value modes  
and the second term contains the residual contributions that are discarded in the original TRG.
Since the diagonal element of $\frac{1}{N_r}{\bm \eta}\cdot{\bm \eta}^\dagger$ is unity for $Z_N$ noise,
we obtain the following expression: 
\begin{align} \label{eq:Mdecompose}
M_{ab} =& \sum_{s=1}^{R} (\sqrt{\Lambda_s} u_{as} \sqrt{\Lambda_s} v_{sb}) 
+ \mathcal{O} \left(\sqrt{\frac{\Lambda_{D_{\rm svd}+1} \Lambda_{D_{\rm svd}+2}}{N_r}} \right).
\end{align}
Thus we can see that all the singular modes are explicitly included in the decomposition as shown in the first term 
and the statistical fluctuation represented in the second term is suppressed by smaller singular values
$\Lambda_{D_{\rm svd}+1}$, $\Lambda_{D_{\rm svd}+2}\ll \Lambda_1$.
This decomposition suggests to modify the third-order tensor $S_{1,2,3,4}$ 
to include the noise vector parts.
For this purpose, we define the following modified third-order tensors as a function of the noise vectors, 
\begin{align}
\label{eq:Sbar}
\bar{S}_{3as}({\bm \eta}) &\equiv 
\begin{cases}
\sqrt{\Lambda_s} u_{as} & (1 \leq s \leq D_{\rm svd}) \\
\sum_{i=D_{\rm svd}+1}^R \sqrt{\frac{\Lambda_i}{N_r}} u_{ai} \eta_{i-D_{\rm svd},s-D_{\rm svd}} & (D_{\rm svd}+1 \leq s \leq D_{\rm svd}+N_r)
\end{cases},
\notag \\
\bar{S}_{1sb}({\bm \eta}) &\equiv 
\begin{cases}
\sqrt{\Lambda_s} v_{sb} & (1 \leq s \leq D_{\rm svd}) \\
\sum_{i=D_{\rm svd}+1}^R \sqrt{\frac{\Lambda_i}{N_r}} \eta_{i-D_{\rm svd},s-D_{\rm svd}}^* v_{ib} & (D_{\rm svd}+1 \leq s \leq D_{\rm svd}+N_r)
\end{cases}, 
\end{align}
and $\bar{S}_{2,4}({\bm \eta})$ are also defined similarly.
Using the modified tensors $\bar{S}_{1,2,3,4}$ and Eq.~\eqref{eq:noise} we immediately see that the original matrix $M$ is recovered
$\displaystyle M_{ab} = \lim_{N_r \to \infty} \sum_{s=1}^{D_{\rm svd} + N_r} \bar{S}_{3as} \bar{S}_{1sb}$ 
for any value of $D_{\rm svd}$. 
We also note that in practice $Z_2$ noise is useful to avoid a complex valued tensor and
we will exclusively use the real valued noise in our calculation given in Sec.~\ref{sec:test}. 

Since the modified third-order tensors contain all the singular modes, 
there is only a statistical error due to the noise vectors instead of the truncation error, 
which ensures that the original matrix should be reproduced up to the statistical error. 
Note that a similar decomposition has already been employed in 
the low-mode approximation for the Dirac propagators in lattice QCD \cite{Foley:2005ac}, 
where the Dirac propagator is dominated by a low-mode part of the Dirac eigenvectors, 
and a high-mode part is stochastically evaluated using the random noise vectors.
The key property of the low-mode approximation of the inverse matrix is reflected in our method.

\section{\label{sec:hybrid}Noise ensemble method}

First we simply use the modified third-order tensor $\bar S$ instead of the original one in the TRG,
however, we do not observe any improvement on the accuracy of the physical quantities, because
the low-rank approximation in the noise method is not as good as the truncated SVD
in the sense of the Frobenius norm 
due to the theorem by Eckhart, Young, and Mirsky.
Thus simply increasing the noise dimension $N_r$ may not give a better accuracy compared to the normal TRG 
when the same number of the bond dimension is used.  
Therefore, instead of increasing $N_r$, we adopt an ensemble approach. 
Namely, we generate an ensemble of random noise vectors 
${\bm\eta}^{[\ell]} \ (\ell=1,\cdots,N)$ with $N$ being the total number of statistics. 
Our strategy is that 
a matrix decomposition is approximately obtained from an ensemble average of the 
modified third-order tensors with keeping a smaller value of $N_r$. 
\begin{align}\label{eq:M}
\displaystyle M_{ab} = \frac{1}{N} \sum_{\ell=1}^N \sum_{s=1}^{D_{\rm svd}+N_r} \bar{S}_{3as} ({\bm \eta}^{[\ell]}) \bar{S}_{1sb} ({\bm \eta}^{[\ell]})
+ \mathcal{O}\left(\sqrt{ \frac{\Lambda_{D_{\rm svd}+1}\Lambda_{D_{\rm svd}+2}}{N_rN} } \right).
\end{align}
Thus, an original matrix can be reproduced in the limit of infinite statistics ($N \to \infty$)
while keeping $N_r$ finite.
Our new stochastic method can be implemented in the TRG, 
and in the next subsections, we discuss how to 
obtain a renormalized tensor 
with two kinds of spatial noise distributions: position-dependent and -independent ways.

\subsection{Coarse graining with position-dependent noise}

\begin{figure}[tbph]
\begin{center}
\includegraphics[clip,width=1\textwidth]{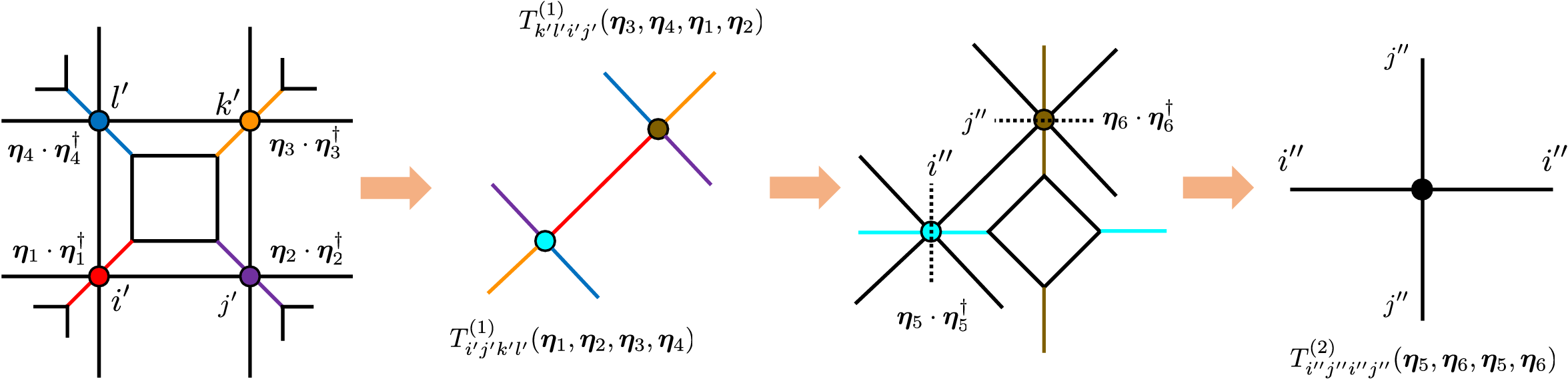} 
\end{center}
\caption{\label{fig:independent} 
Coarse-graining of the tensor network using the position-dependent noise
for $2\times 2$ lattice with the periodic boundary condition for all directions.
In the initial lattice four independent noise vectors ${\bm \eta}_{1,2,3,4}$ are 
distributed to the four sites, and the initial tensors are decomposed.
The contraction for each plaquette produces a renormalized tensor $T^{(1)}$ on $1\times 2$ lattice.
And then two different noise vectors ${\bm \eta}_{5,6}$ are distributed and the tensors are decomposed.
Finally after contracting for one plaquette, a trace of a renormalized tensor $T^{(2)}$ is obtained.}
\end{figure}

The first step is to 
decompose the original tensor as in Eq.\eqref{eq:M} and obtain the third-order tensors $\bar S_{1,2,3,4}$.
Since the decomposition is separately done for each lattice site $i$, the noise vector is position dependent ${\bm \eta}^{[\ell]}_i$.
Using $\bar S$ 
we define a renormalized tensor $T^{(1)[\ell]}_{i'j'k'l'}({\bm \eta}_1,{\bm \eta}_2,{\bm \eta}_3,{\bm \eta}_4)$ as   
\begin{align}
\label{eq:contg}
T^{(1)[\ell]}_{i'j'k'l'}({\bm \eta}_1,{\bm \eta}_2,{\bm \eta}_3,{\bm \eta}_4) = {\rm Tr}[ \bar{S}_{1i'}({\bm \eta}^{[\ell]}_1) \bar{S}_{2j'}({\bm \eta}^{[\ell]}_2) 
\bar{S}_{3k'}({\bm \eta}^{[\ell]}_3) \bar{S}_{4l'}({\bm \eta}^{[\ell]}_4)], 
\end{align}
where the tensor indices $i', j', k'$, and $l'$ run from 1 to $D_{\rm svd}+N_r$,
and hence the bond dimension $D_{\rm cut}$ for a renormalized tensor $T^{(1)}$ is given as $D_{\rm cut}=D_{\rm svd}+N_r$.
By iterating this RG process $n$ times, 
we obtain a sequence of renormalized tensors for each sample $\ell$, 
\begin{align}
T^{(0)} \to T^{(1)[\ell]} \to T^{(2)[\ell]} \cdots \to T^{(n)[\ell]}, \ \ (\ell=1, \cdots, N).
\end{align}
Since the sets of random noise vectors are generated 
for each RG step and for each site independently,  
the noise vectors satisfy the following property in orthogonality, 
\begin{align} 
\label{eq:ortho1}
\frac{1}{N}\sum_{\ell=1}^N  \frac{1}{N_r} \bm{\eta}^{[\ell]}_{i} \cdot \bm{\eta}^{[\ell]\dagger}_{j} 
&= 
\delta_{i,j} {\bf 1}_{D \times D} +\mathcal{O}(1/\sqrt{N_rN}),
\end{align}
and this orthogonal relation can be generalized to arbitrary order tensor products. 
For example, 
the tensor product of two independent noise vectors of $\bm{\eta}_i^{[\ell]}$ 
and $\bm{\eta}_j^{[\ell]}$ for different sites $i \neq j$ should also have an orthogonality 
\begin{align} \label{eq:ortho2}
\frac{1}{N}\sum_{\ell = 1}^N 
\frac{1}{N_r} \left( \bm{\eta}^{[\ell]}_{i} \cdot \bm{\eta}^{[\ell]\dagger}_{i}\right)_{ab}
& \otimes \frac{1}{N_r} \left( \bm{\eta}^{[\ell]}_{j} \cdot \bm{\eta}^{[\ell]\dagger}_{j}\right)_{cd} 
\\ \notag 
= ({\bf 1}_{D \times D})_{ab} & \otimes ({\bf 1}_{D \times D})_{cd} +\mathcal{O}(1/\sqrt{N_rN}) \quad ({\rm{for}}\  i \neq j).
\end{align}
Thanks to these properties,  
the truncation error in the tensor decomposition is completely replaced with the statistical error of the random noise. 
The exact partition function $Z_V$ for a finite volume $V=2^n$ 
can be obtained from the renormalized tensor $T^{(n)[\ell]}$ in the limit of infinite statistics
\begin{align}\label{eq:Zv}
Z_V = \lim_{N\to\infty} \frac{1}{N} \sum_{\ell=1}^N Z(T^{(n)[\ell]}), 
\end{align}
where $Z(T^{(n)[\ell]})$ is a sample of the partition function, 
\begin{align}
Z(T^{(n)[\ell]}) =& {\rm Tr}\left[T^{(n)[\ell]}\right] = \sum_{i,j=1}^{D_{\rm{cut}}} T^{(n)[\ell]}_{ijij}.
\end{align}
See Fig.~\ref{fig:independent} for multiple coarse-graining procedures in this method.
%
%

At each coarse-graining step, we obtain an ensemble of tensor configurations 
so that this method is similar to a Monte Carlo method for configuration generations, 
while in our case the tensor variables are stochastically generated 
in analogous to a random walk and hence there is no sign problem 
and the individual configurations are completely uncorrelated. 
Thus individual configurations can be efficiently generated by using massive parallel computers.
Furthermore since the contribution of all modes of SVD are maintained in the coarse-graining step,
it is mathematically guaranteed that an exact result should be obtained within the statistical error.
On the other hand, the lattice homogeneity is completely broken due to the position-dependent noise, 
thus we need to separately calculate a renormalized tensor for each plaquette, 
where we also have to take into account the boundary conditions in a finite volume.
Therefore, the system volume $V$ has to be fixed beforehand,
which restricts on the maximum number of the coarse-graining processes.
If we repeat the processes until obtaining a single tensor, 
the total computational cost will be as expensive as $\mathcal{O}(N V D_{\rm cut}^6)$, 
so that it spoils one of advantages in the original TRG methods: a logarithmic computational cost on the volume.
All the above properties are already pointed out in the previous studies of the MCTN
\cite{ferris2015unbiased, 2017arXiv171003757H}.
The difference between the MCTN and our position-dependent noise method lies in the probability distribution of which modes are taken in.
Thanks to the random noise vectors, the renormalized tensors are compact but contain all singular modes 
(all-mode renormalization). 
We consider that our case is rather simple to implement and can easily control the statistical error.
In the next section we shall numerically study the position-dependent method in details.

\subsection{Coarse graining with common noise}
\label{sec:coarse_graining_common_noise}
\begin{figure}[tbph]
\begin{center}
\includegraphics[clip,width=0.7\textwidth]{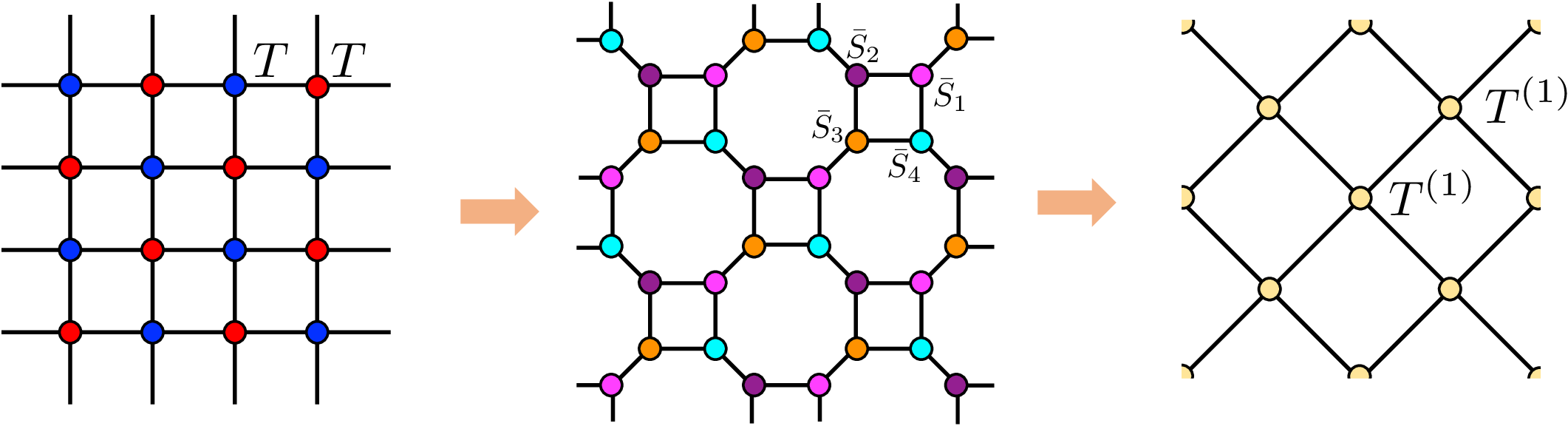} 
\end{center}
\caption{\label{fig:common} 
A coarse-graining of the tensor network with common noises.
In left panel, an initial tensor $T$ is uniformly put on each site, 
while even and odd sites are distinguished by the different sets of the noise vectors 
represented by two different colors on sites, that is,
the common noise ${\bm\eta}_1^{[\ell]}$ (${\bm\eta}_2^{[\ell]}$ ) is used on all red (blue) sites.
In middle panel, the third-order tensors $\bar{S}_{1,2,3,4}$ are obtained by a tensor decomposition.
Right panel shows that after coarse graining a renormalized tensor $T^{(1)}$ is commonly obtained for each site,
namely we obtain a uniform tensor network consisting of $T^{(1)}$.}
\end{figure}

In order to preserve the lattice homogeneity
we consider a common (position-independent) noise.
For this purpose we use a common set of noise vectors 
to obtain $\bar{S}_{1,3}$ or $\bar{S}_{2,4}$,
\begin{align}
T_{ijkl}=\displaystyle M_{ab} \simeq \frac{1}{N} \sum_{\ell=1}^N \sum_{s=1}^{D_{\rm svd}+N_r} \bar{S}_{3as} ({\bm \eta_1}^{[\ell]}) \bar{S}_{1sb} ({\bm \eta_1}^{[\ell]}),
\\
T_{ijkl}=\displaystyle \tilde M_{cd} \simeq \frac{1}{N} \sum_{\ell=1}^N \sum_{s=1}^{D_{\rm svd}+N_r} \bar{S}_{4cs} ({\bm \eta_2}^{[\ell]}) \bar{S}_{2sd} ({\bm \eta_2}^{[\ell]}),
\end{align}
where two sets of noise vectors ${\bm \eta}^{[\ell]}_1$ and ${\bm \eta}^{[\ell]}_2$ are independent with each other in general (see Fig.~\ref{fig:common}).
The renormalized tensors are then obtained as 
\begin{align}
\label{eq:renormalization}
T^{(1)[\ell]}_{i'j'k'l'}({\bm \eta}_1,{\bm \eta}_2,{\bm \eta}_1,{\bm \eta}_2) = {\rm Tr}[ \bar{S}_{1i'}({\bm \eta}^{[\ell]}_1) \bar{S}_{2j'}({\bm \eta}^{[\ell]}_2) 
\bar{S}_{3k'}({\bm \eta}^{[\ell]}_1) \bar{S}_{4l'}({\bm \eta}^{[\ell]}_2)].
\end{align}
Since in this method the same renormalized tensor is obtained at every site after a coarse graining, 
we only need to calculate a tensor contraction at a single plaquette for each RG step. 
Thus the order of the computational cost per sample remains the same as that of the original TRG. 
We refer this method as common noise method on normal lattice.

We note, however, that there exists an unwanted additional systematic error 
that arises from a contact term of the same noise vectors. 
Let $\bm{\eta}$ be a noise vector,  
which satisfies the relation
\begin{align} 
\frac{1}{N_r} \left( \bm{\eta} \cdot \bm{\eta}^\dagger \right)_{ab}
= \frac{1}{N_r} \sum_{r}^{N_r} \eta_{ar}\eta_{br}^*
= \delta_{ab} + \mathcal{O}(1/\sqrt{N_r}).
\end{align}
Since we have to commonly use $\bm{\eta}$
for tensor decompositions at two different sites, 
e.g.,  see the argument of $\bar{S}_{1}$ and $\bar{S}_{3}$ in Eq.~\eqref{eq:renormalization}, 
we encounter the following tensor product in the tensor contraction for a single 
plaquette as in Eq.~\eqref{eq:renormalization}, 
\begin{align} \notag
\frac{1}{N_r^2} \left( \bm{\eta} \cdot \bm{\eta}^{\dagger} \right)_{ab}
\otimes \left( \bm{\eta} \cdot \bm{\eta}^{\dagger} \right)_{cd} 
=& \frac{1}{N_r^2} 
\sum_{r_1=1}^{N_r} \eta_{ar_1} \eta_{br_1}^* 
\sum_{r_2=1}^{N_r} \eta_{cr_2} \eta_{dr_2}^*
\\ \notag 
=& 
\frac{N_r-1}{N_r^2}
\sum_{r_1=1}^{N_r} \eta_{ar_1} \eta_{br_1}^* 
\left( \delta_{cd}+\mathcal{O}(1/\sqrt{N_r}) \right)
\\ \notag
& 
+ 
\frac{1}{N_r^2}
\sum_{r_1=1}^{N_r} \eta_{ar_1} \eta_{br_1}^* 
\eta_{cr_1} \eta_{dr_1}^*
+ \mathcal{O}(1/\sqrt{N_r})
\\ \label{eq:contact}
=& 
\delta_{ab}\delta_{cd}+\frac{1}{N_r} \delta_{ad}\delta_{cb}(1-\delta_{ab}\delta_{cd}) + \mathcal{O}(1/\sqrt{N_r}), 
\end{align}
where the second term is the contact term due to a multiple use of noise vectors,
which causes a noise cross contamination effect and 
violates the orthogonal relation.\footnote{We implicitly assume $Z_N$ random noises with $N \geq 3$ in Eq.~\eqref{eq:contact}. 
In the case of $Z_2$ noise, there may exist additional contact terms of order $1/N_r$, since two vectors of $\eta_r$ and $\eta_r^*$ can not be distinguished.} 
In terms of the tensor networks 
this contamination would generate an unphysical network that links between two distant tensors. 
Nevertheless, from Eq.~\eqref{eq:contact}, 
we find that this residual systematic error is proportional to $1/N_r$,
thus it is reduced for large $N_r$. 
Moreover, thanks to its simple functional form in terms of $N_r$,
the systematic error in the free energy can be controlled in a straightforward way. (See Fig.~\ref{fig:D20_S20}
for $1/N_r$ scaling in actual simulation result of the common noise method in the next section.)

\begin{figure}[tbph]
\begin{center}
\includegraphics[clip,width=0.8\textwidth]{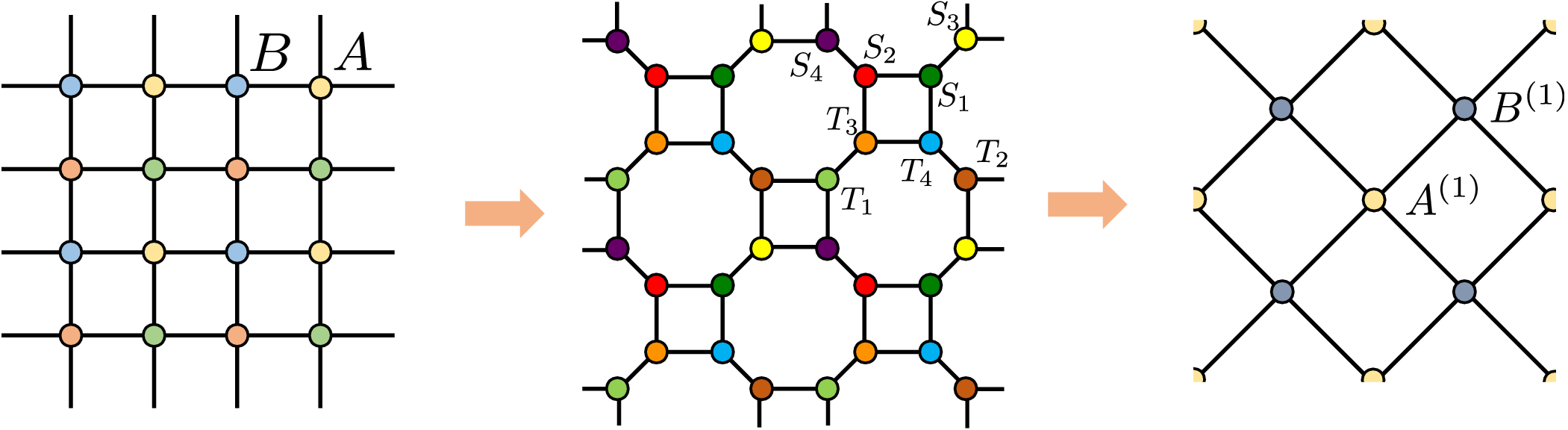} 
\end{center}
\caption{\label{fig:checker-board} 
Coarse graining of the tensor network with common noises for checkerboard lattice.
}
\end{figure}

To reduce the noise correlation, equivalently, systematic error of the common noise method
we consider a checkerboard lattice, 
where four different sets of the noise vectors are distributed at four sites on a plaquette, respectively (see Fig.~\ref{fig:checker-board}).
Suppose that in the initial tensor network two different tensors $A$ and $B$ are on even and odd sites, respectively.  
We decompose these two tensors 
with four different noise vectors distributed at four sites 
so that we obtain eight different third-order tensors $T_{1,2,3,4}$ and $S_{1,2,3,4}$. 
After contracting these tensors 
we obtain two different renormalized tensors $A^{(1)}$, and $B^{(2)}$ defined as  
\begin{align}
\label{eq:contAB}
A^{(1)} = {\rm Tr}[ T_1 T_2 S_3 S_4], 
\quad \quad 
B^{(1)} = {\rm Tr}[ S_1 S_2 T_3 T_4].
\end{align}
As shown in Fig.~\ref{fig:checker-board}, 
after the coarse-graining 
the same checkerboard lattice structure appears, 
where the two different tensors $A^{(1)}$ and $B^{(1)}$ 
reside on even and odd sites, respectively.
Thus the iterative coarse-graining 
process is possible. 
The computational cost in the checkerboard lattice case is doubled compared to the normal lattice case, 
since both the tensor decompositions and contractions should be performed for even and odd sites separately. 
We note that although the noise correlation of the checkerboard lattice is reduced compared with the normal lattice,
there still exists the contamination error 
since two distant noise vectors come into contact after multi coarse-graining processes. 
After all, the existence of a contact term is inevitable to preserve the lattice homogeneity, 
and we end up with taking the residual systematic error into account in the common noise methods. 
We will discuss a detailed study of the systematic error and its scaling property in the next section.

\section{\label{sec:test}Numerical test in 2D Ising model}

We implement the noise methods in the TRG for Ising model on square lattice
and study the efficiency of 
the position-dependent and the common noise methods
in comparison with the original TRG as well as the analytic results on finite volumes~\cite{Kaufman:1949ks} 
and the Onsager's exact result in infinite volume limit. 
We employ the $Z_2$ noise, i.e., $\eta_{ir}$ takes $\pm 1$ randomly 
and calculate the free energy density with volume $V=2^n$. 
In what follows, we refer to $n$ as the number of the coarse-graining steps, 
$N$ the total number of statistics, 
and $D_{\rm cut}=D_{\rm svd}+ N_r$ the total number of the bond dimensions, 
where $D_{\rm svd}$ and $N_r$ are the bond dimensions for
the singular modes included exactly and the dimension of the noise vectors, respectively.

\subsection{Position-dependent noise}
\label{sec:position_dependent_noise}
We first examine the position-dependent noise method.
As shown in Eq.~\eqref{eq:Zv},  
there is no systematic error even for a finite $D_{\rm cut}$ in this method, 
so an exact partition function $Z_V$ on finite volume $V=2^n$ 
is obtained as an ensemble average of the renormalized tensor $T^{(n)}$ at $n$th coarse-graining step. 
The free energy density $f_V$ is calculated as 
\begin{align}\label{eq:fV}
f_V =& -\frac{T}{V}\log{Z_V}, 
\end{align}
where $Z_V$ is the mean value of the partition function given in Eq.~\eqref{eq:Zv}.
To estimate a statistical error we utilize the jackknife method, 
by which the statistical fluctuation of the logarithmic function 
is easily estimated. 
We note, however, that in the stochastic method 
a negative $Z$ sample $[Z(T^{(n)[\ell]})<0]$ can be occasionally generated, 
and some of the jackknife samples also become negative due to a limited number of statistics. 
Since we have to exclude negative jackknife samples from the error estimate, 
the statistical error of $f_V$ is not estimated correctly.  
We will mention it when we present such a result. 
On the other hand, to quote the central value for $f_V$
we take the ensemble average of both positive and negative $Z$ samples. 
An implication of negative $Z$ contributions is discussed in the following analysis.

\begin{figure}[tbph]
\begin{center}
\includegraphics[clip,width=0.5\textwidth]{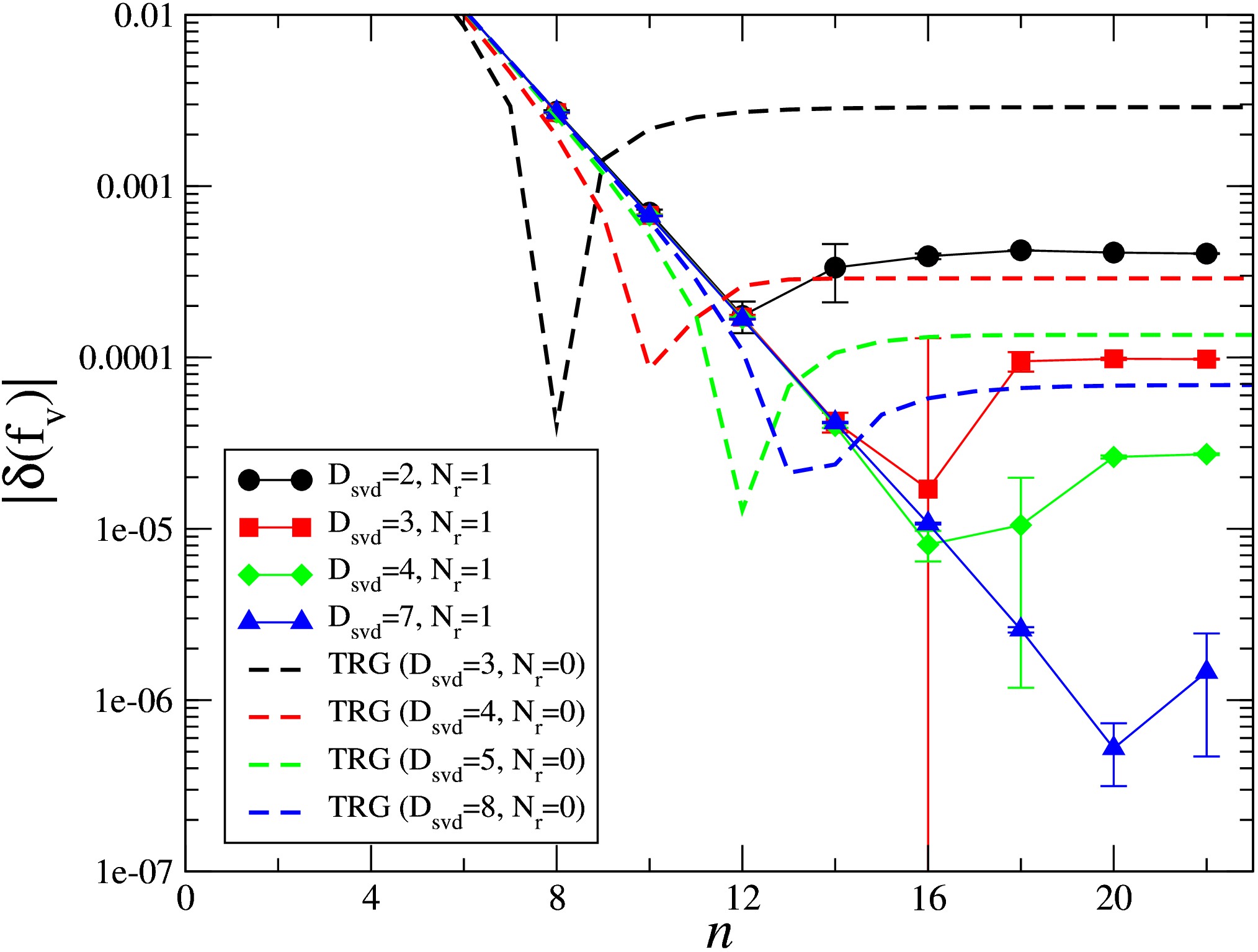} 
\end{center}
\caption{\label{fig:independent_N500} 
The absolute value of the relative error $|\delta(f_V)|$ at $T=T_{\rm c}$ as a function of
the coarse-graining step.
The data points represent the results for the position-dependent noise method for $N=500$.
The dashed curves show the results obtained from the TRG method 
with the bond dimension $D_{\rm cut}=D_{\rm svd}+N_r$,
e.g., the black dashed line represents the TRG results with $D_{\rm cut}=3$.
Note that there are negative jackknife samples for $n \geq 16$ that are excluded from 
the jackknife error estimate.  
}
\end{figure}

\begin{figure}[tbph]
\begin{center}
\includegraphics[clip,width=0.3\textheight]{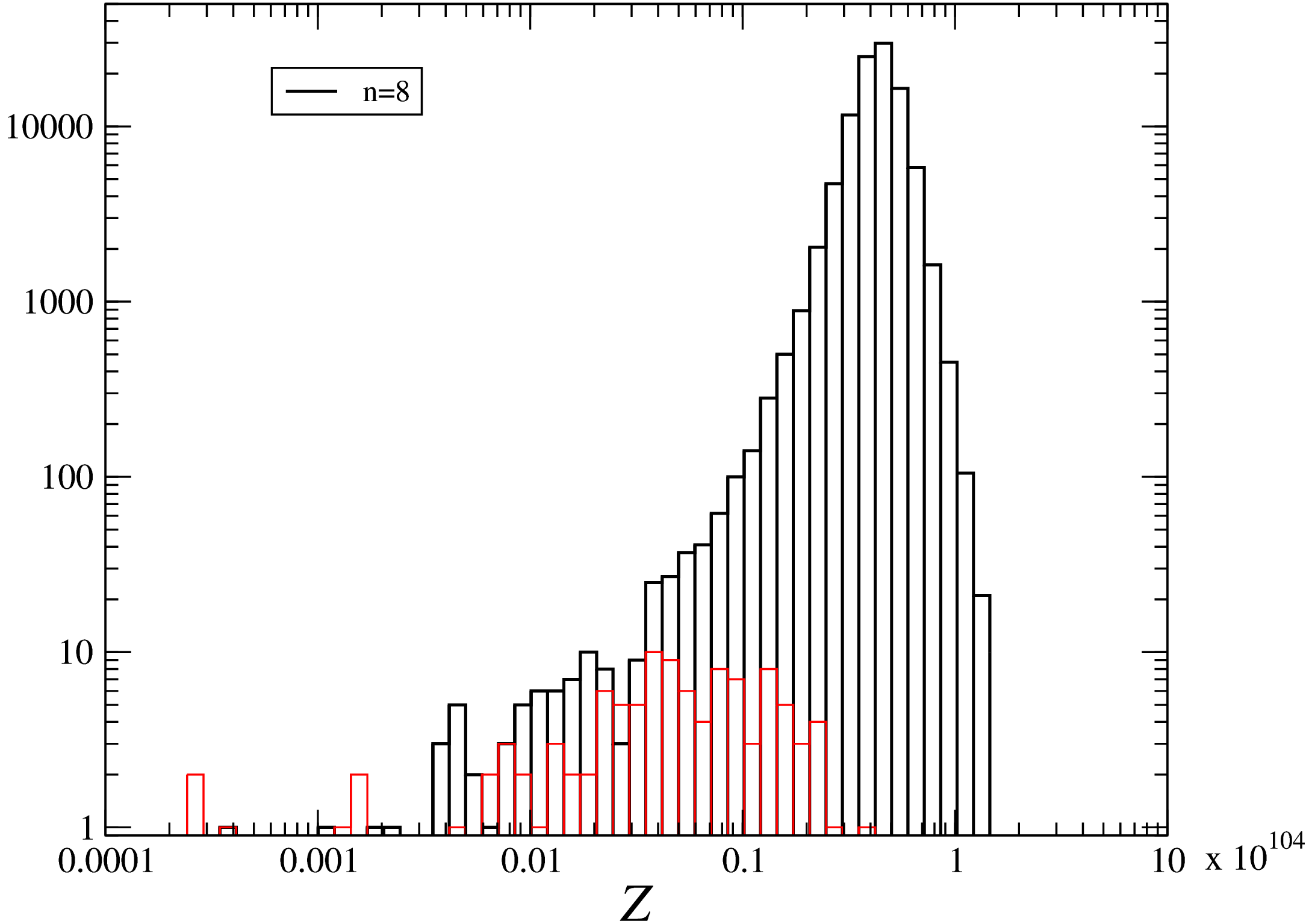} 
\quad 
\includegraphics[clip,width=0.3\textheight]{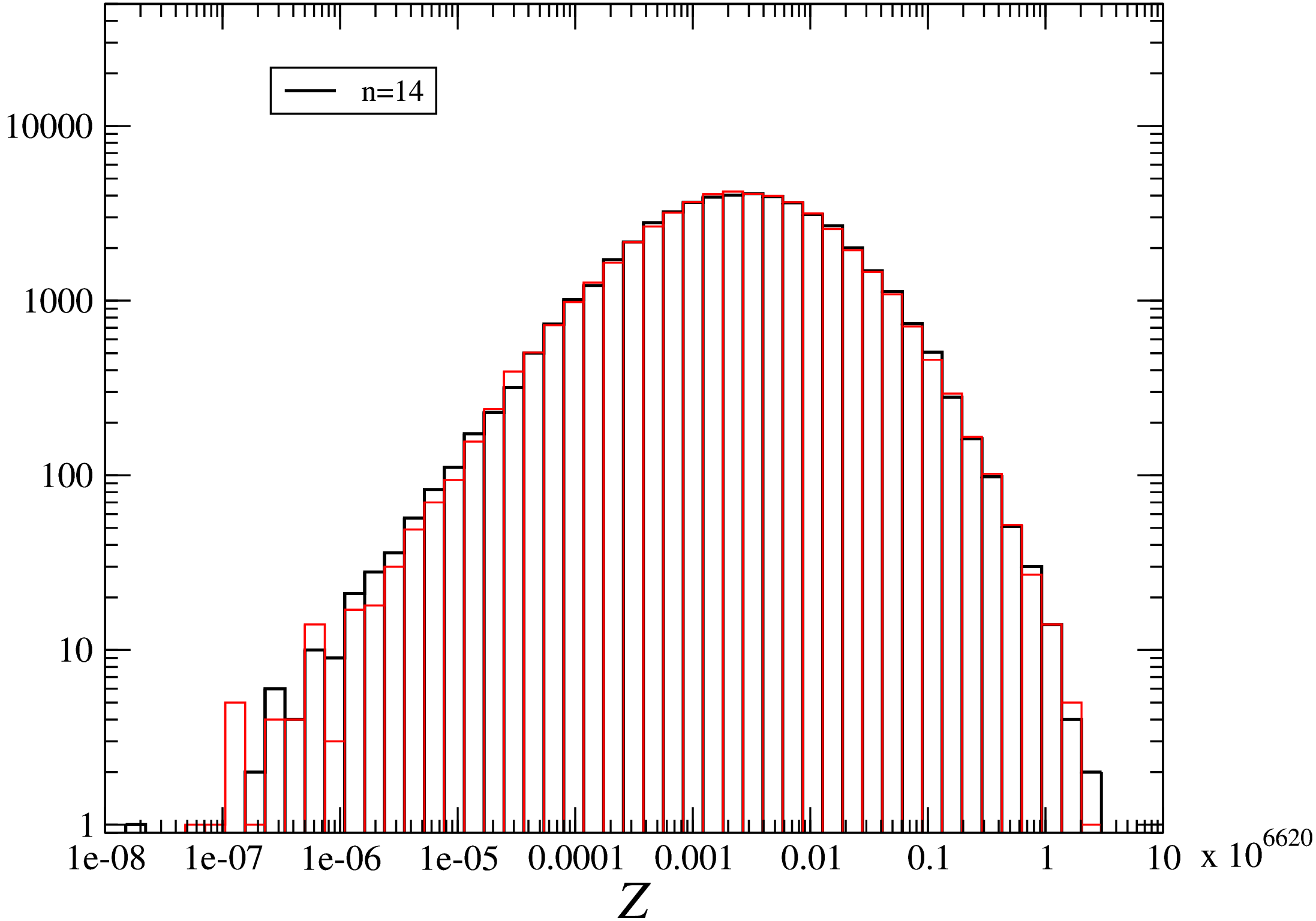} 
\end{center}
\caption{\label{fig:histogram} 
Histogram of $Z$ for $V=2^8$ (left) and $V=2^{14}$ (right)
for 100000 samples with $D_{\rm svd}=2$ and $N_r=1$ at $T=T_{\rm c}$ ,
where the numbers of positive and negative $Z$ samples are indicated as black and red bars, respectively.
}
\end{figure}

\begin{figure}[tbph]
\begin{center}
\includegraphics[clip,width=0.45\textwidth]{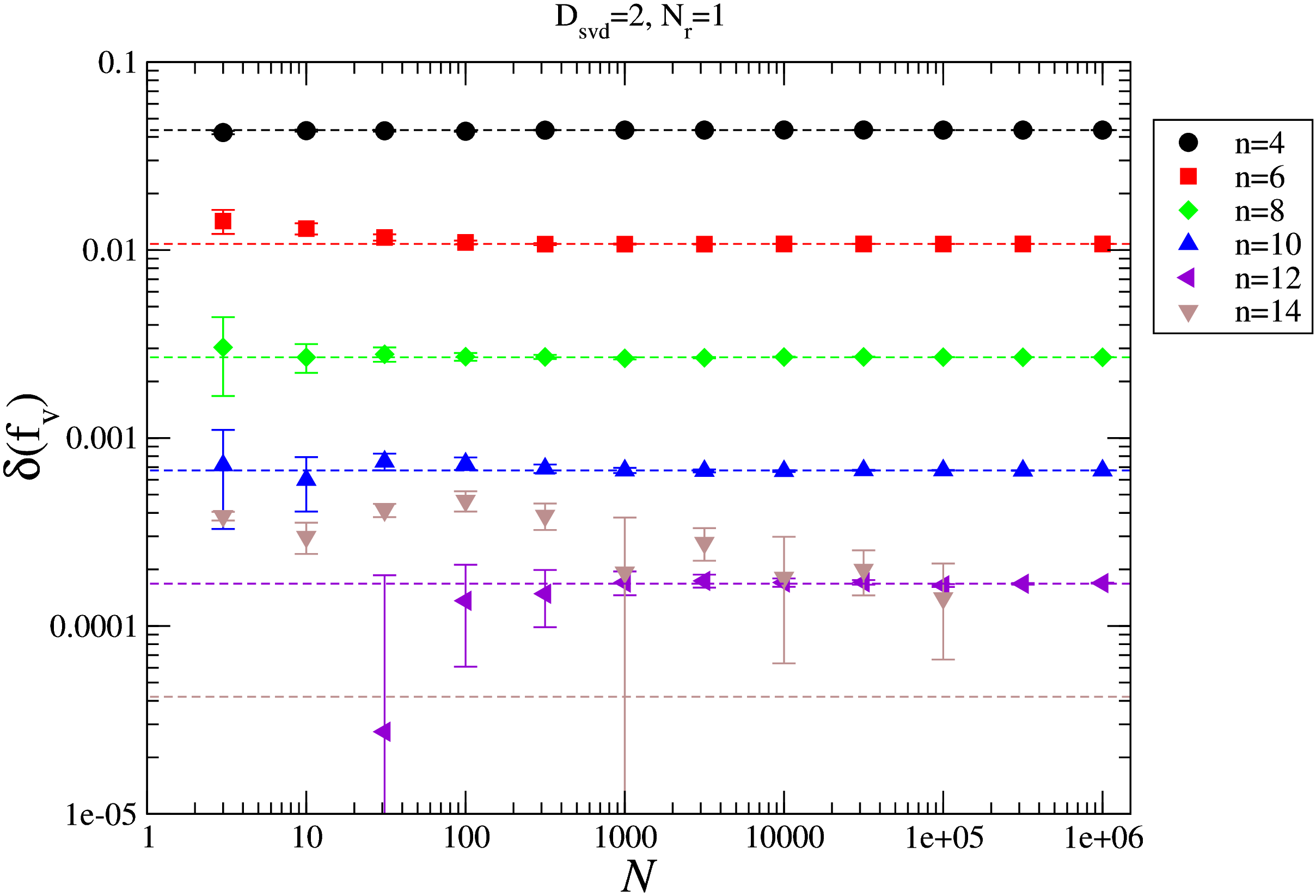} 
\quad 
\includegraphics[clip,width=0.45\textwidth]{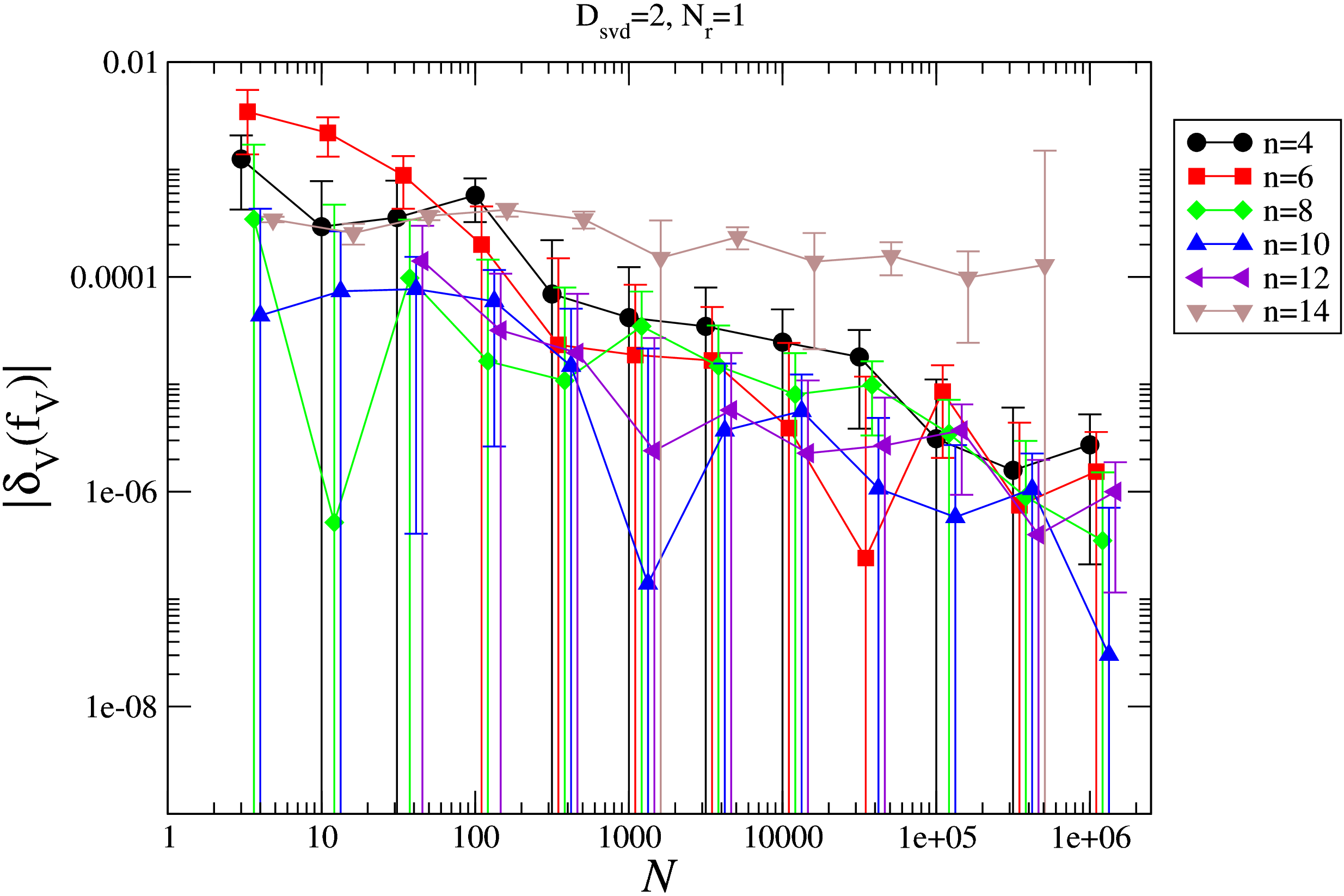} 
\end{center}
\caption{\label{fig:N-dependence} 
Sample size dependence for  $\delta (f_V)$ (left panel) 
and $|\delta_V(f_V)|$ (right panel) at $T=T_{\rm c}$ with $D_{\rm svd}=2, N_r=1$
for various volumes of $V=2^n$ with $n=4, 6, 8, 10, 12,$ and 14. 
The horizontal dashed lines show 
the analytic results on finite volumes~\cite{Kaufman:1949ks}. 
In the right panel, the values are slightly shifted along the $x$ axis for clarity. 
Note that there are negative jackknife samples in $N=10^3$ and $N=10^{5.5}$ statistics at $n = 14$ 
that are excluded from the jackknife error estimate.   
}
\end{figure}

Figure \ref{fig:independent_N500} 
shows the absolute value of relative error of the free energy $|\delta(f)|$,
where $\delta(f)=(f-f^{\rm exact})/f^{\rm exact}$ is a relative error from the Onsager's exact result $f^{\rm exact}$ 
in the infinite volume.  
In the figure, we show the results for several values of $D_{\rm svd}$ with fixed $N_r=1$ at the critical temperature $T=T_{\rm c}$. 
We use 
$N=500$ statistics for each parameter, 
where we exclude some negative jackknife samples from the error analysis. 
As shown in the figure, 
we find that our results with $\mathcal{O}(100)$ statistics basically
give a much more accurate result than the TRG for each coarse-graining step. 
We also observe that 
the results tend to be flat as increasing the coarse-graining step $n$.
This flattening behavior apparently contradicts what explained in the previous section, that is,
the results of the position-dependent noise method should not have systematic errors so as to be consistent with the exact result. 
A possible reason for the behavior is a lack of statistics and the error is not correctly estimated.
In fact, when computing a partition function itself using a statistical method,
one needs extremely high statistics especially for large volume case since
the partition function has a broad distribution whose width grows exponentially with respect to a system volume 
(see, e.g., \cite{Plefka:1996tz,ferris2015unbiased}). 
Such a distribution of the partition function can be understood as follows.
Since we approximate the exact partition function $Z_V^{\rm ex}$ 
as tensor products 
\begin{align} 
Z_V^{\rm ex} = {\rm Tr}\left[ \prod_{i=1}^{2^n}\otimes T^{(0)} \right] 
\simeq 
\frac{1}{N}\sum_{\ell=1}^N {\rm Tr}\left[ \prod_{i=1}^{2^{n-1}}\otimes T^{(1)[\ell]} \right] 
\simeq \cdots \simeq 
\frac{1}{N}\sum_{\ell=1}^N {\rm Tr}\left[T^{(n)[\ell]}\right],  
\end{align}
where each $T^{(n)[\ell]}$ behaves like a random variable due to noise vectors, 
the partition function 
given as a product of the random tensors 
is expected to have a log-normal distribution. 
To see if the situation holds true, in Fig.~\ref{fig:histogram},
we plot the histograms of $Z(T^{(n)[\ell]})$ 
whose distribution is well described by a log-normal distribution, 
and central value is proportional to the exponential of the volume $V$, thus the distribution of $Z$ is confirmed to follow the log-normal one.
In addition, we also find that a probability of negative samples 
increase as volume ($n$) increases, 
and the total amount of the negative $Z$ contributions will become 
almost the same as that of the positive $Z$ contributions. 
The result at $n=14$ actually is the case.
In this case it is notoriously difficult to numerically calculate any thermodynamical quantity based on the stochastic approach. 
To directly see this, we show the sample size dependence of $\delta (f_V)$ and $|\delta_V(f_V)|$
in Fig.~\ref{fig:N-dependence}, 
where $\delta_V(f_V)=(f_V-f_V^{\rm analytic})/f_V^{\rm analytic}$ is a relative error from the analytic result on finite volume ($f_V^{\rm analytic}$)~\cite{Kaufman:1949ks}.
In the case of $n \leq 12$, 
the results with sufficient statistics ($N \gtrsim 100$) are consistent with $f_V^{\rm analytic}$ 
within the statistical error and become more precise with more statistics. 
These trends are clearly observed on the right panel, where $|\delta_V(f_V)|$ with $n \leq 12$ is consistent with zero
and its central value and statistical error decrease with increasing $N$.
We thus numerically confirm that the position-dependent noise method 
provides an unbiased result for $n \leq 12$ as it should be. 
However, in the case of $n=14$, 
the mean value $f_V$ becomes unstable and it is difficult to obtain the $1/\sqrt{N}$ scaling of the error. 
In order to alleviate this numerical difficulty in large $n$ region $D_{\rm cut}$ should be increased.

\begin{figure}[htbp]
\begin{center}
\includegraphics[clip,width=0.3\textwidth]{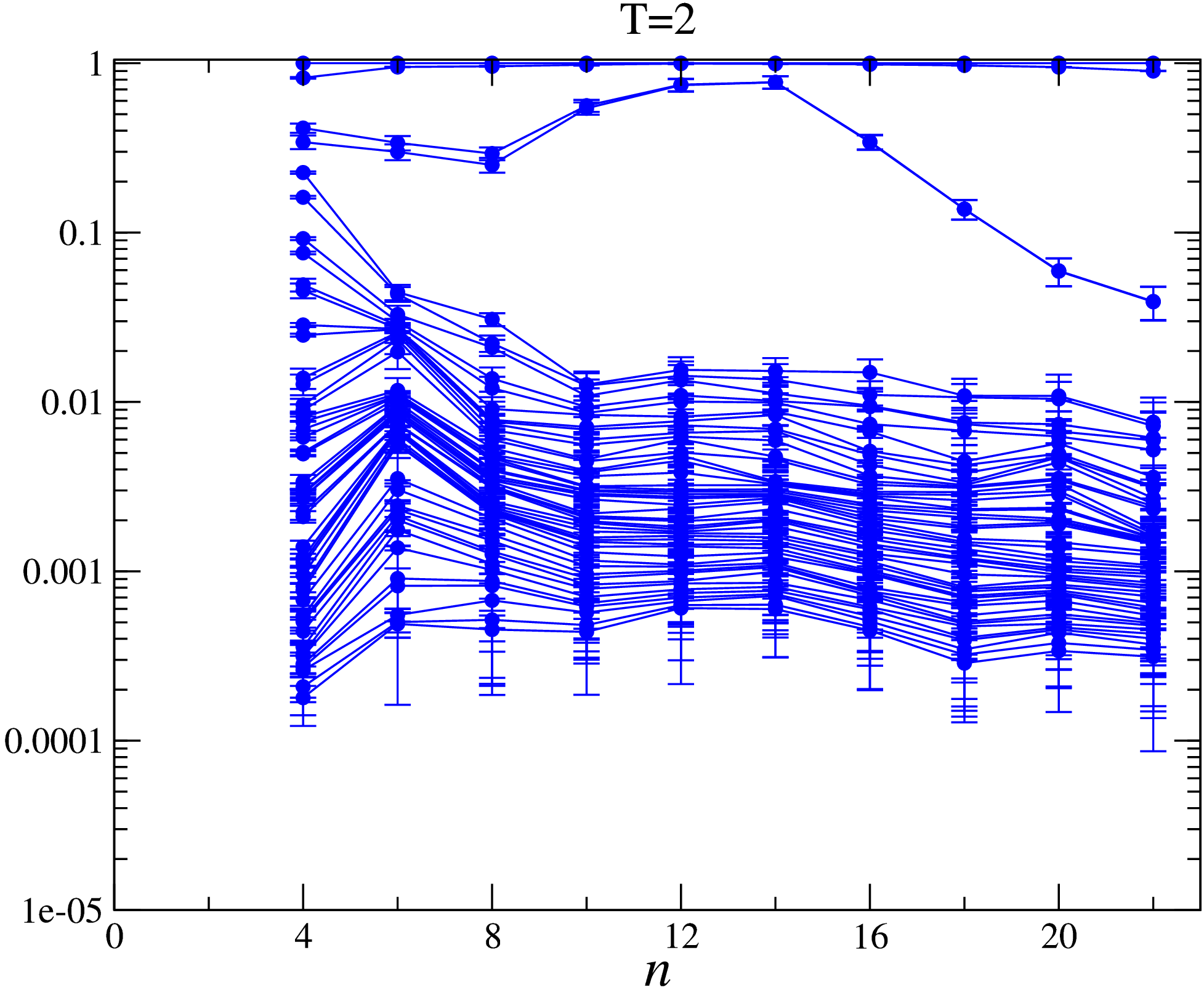} 
\includegraphics[clip,width=0.3\textwidth]{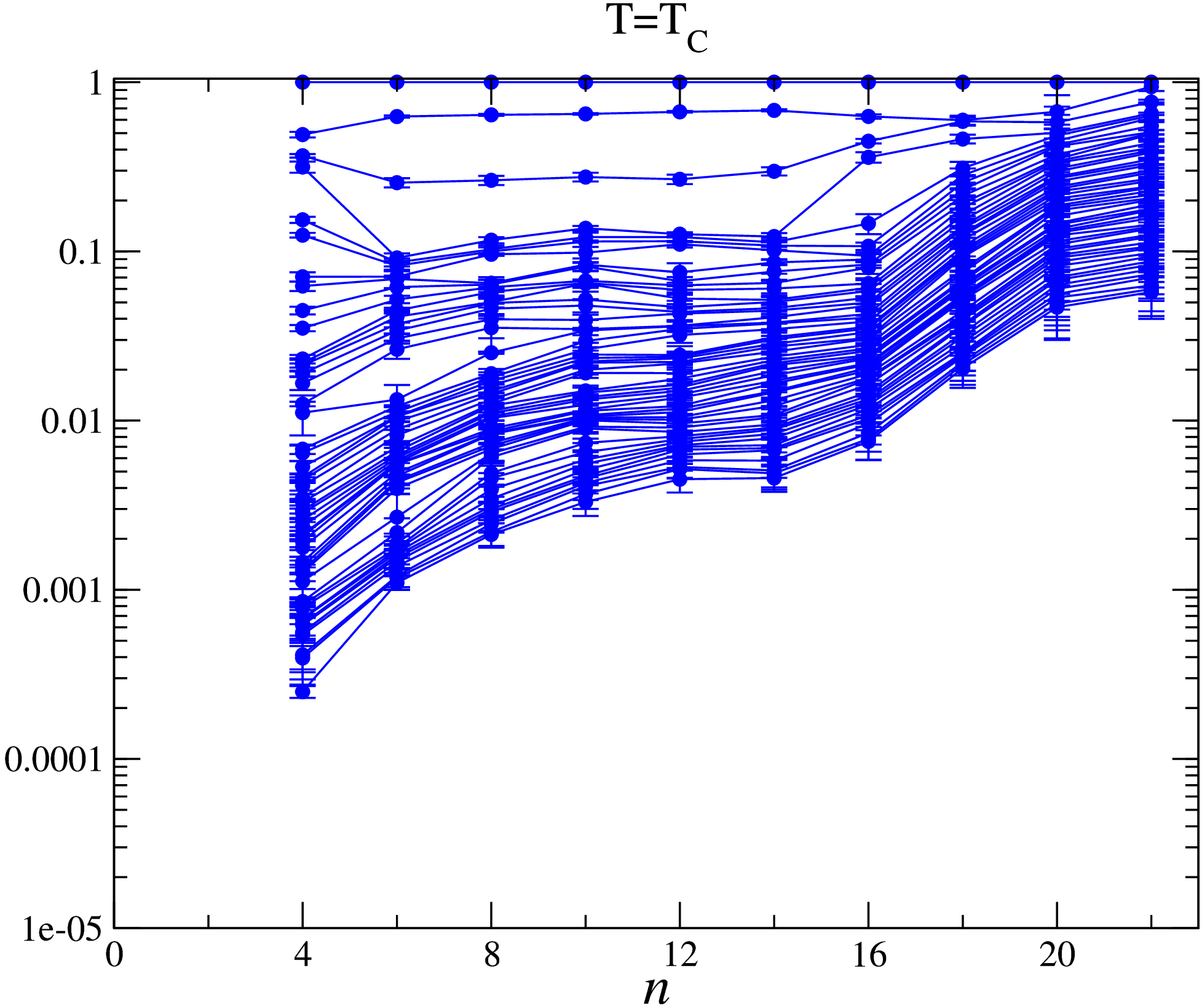} 
\includegraphics[clip,width=0.3\textwidth]{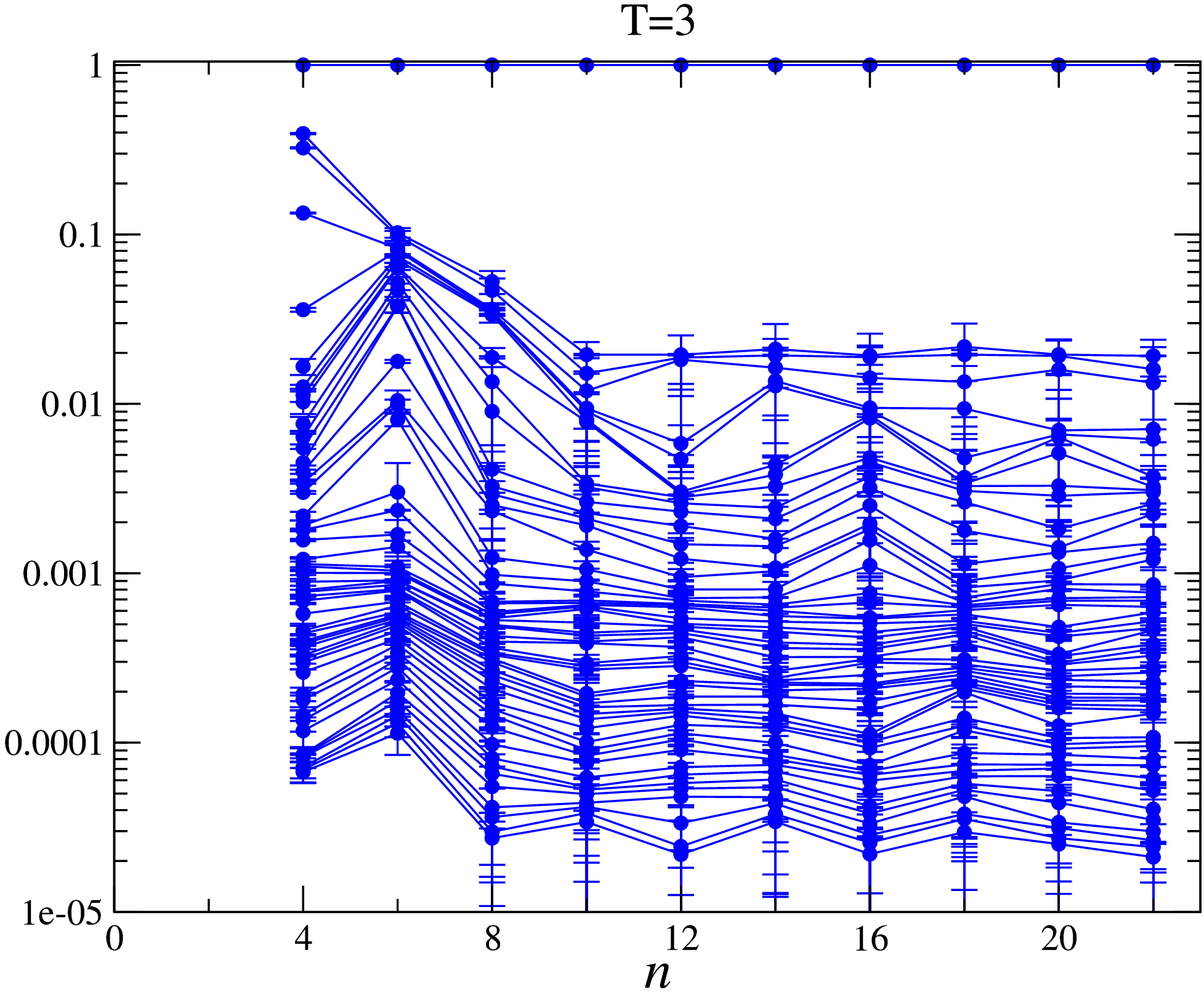} 
\end{center}
\begin{center}
\includegraphics[clip,width=0.3\textwidth]{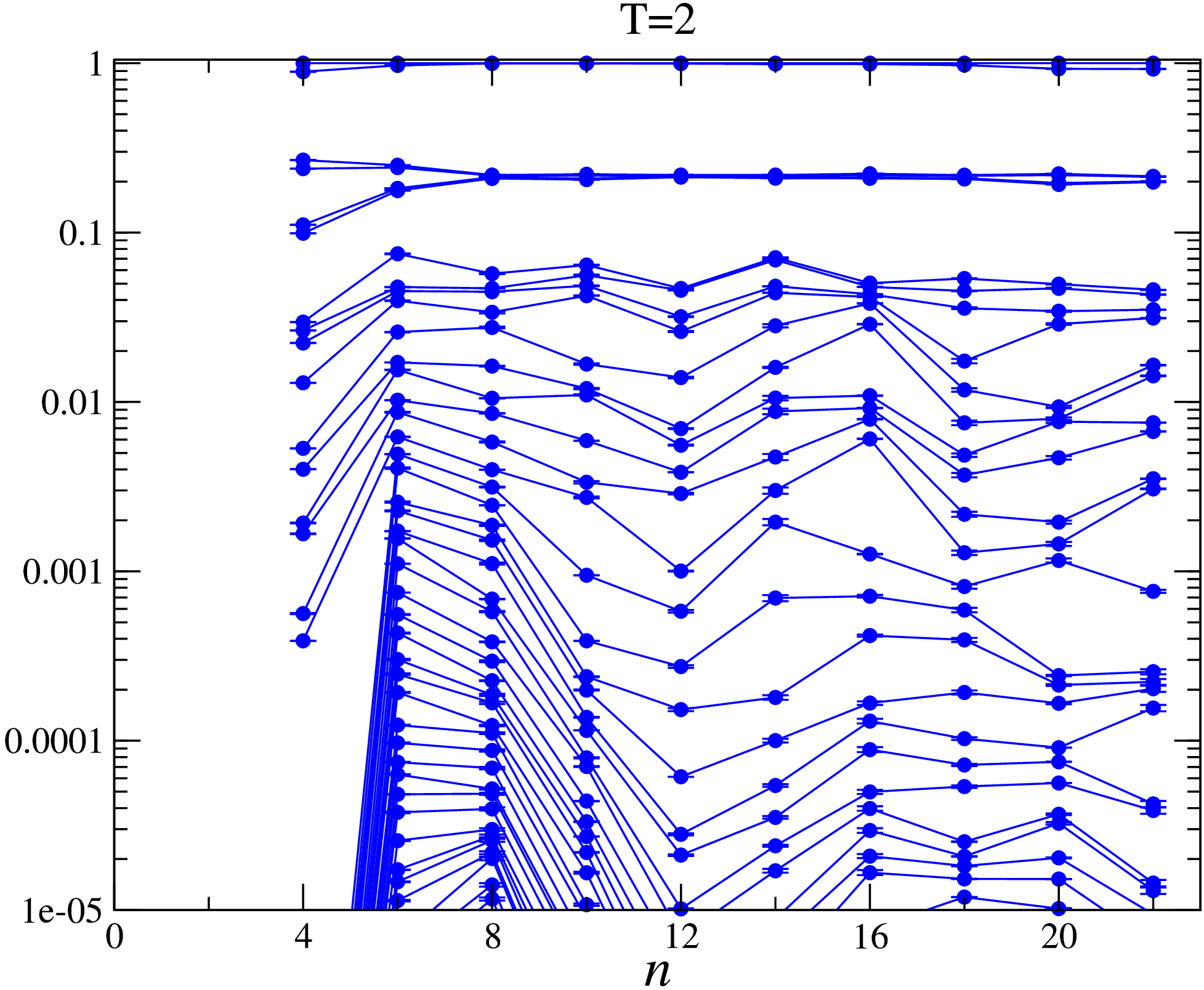} 
\includegraphics[clip,width=0.3\textwidth]{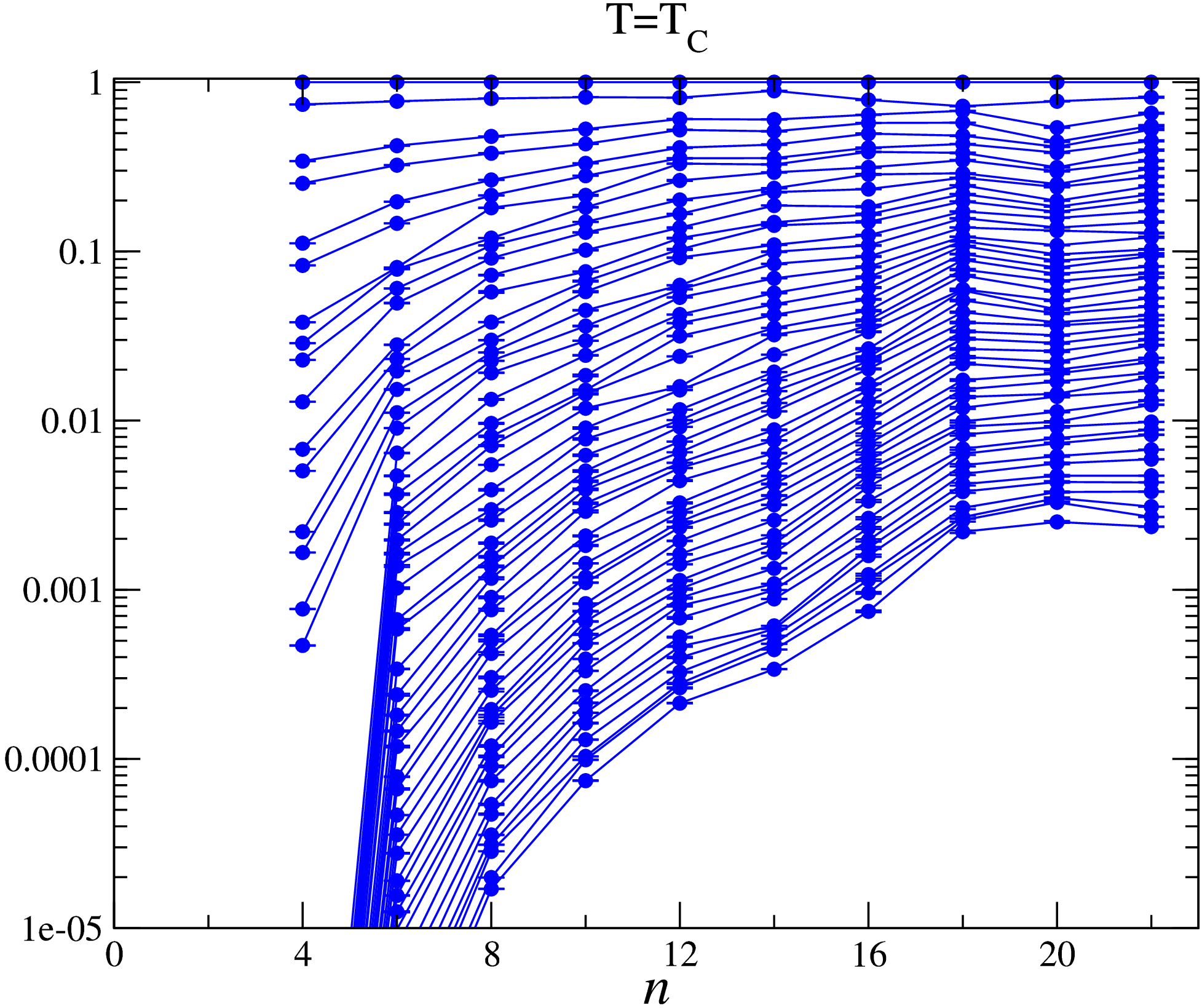} 
\includegraphics[clip,width=0.3\textwidth]{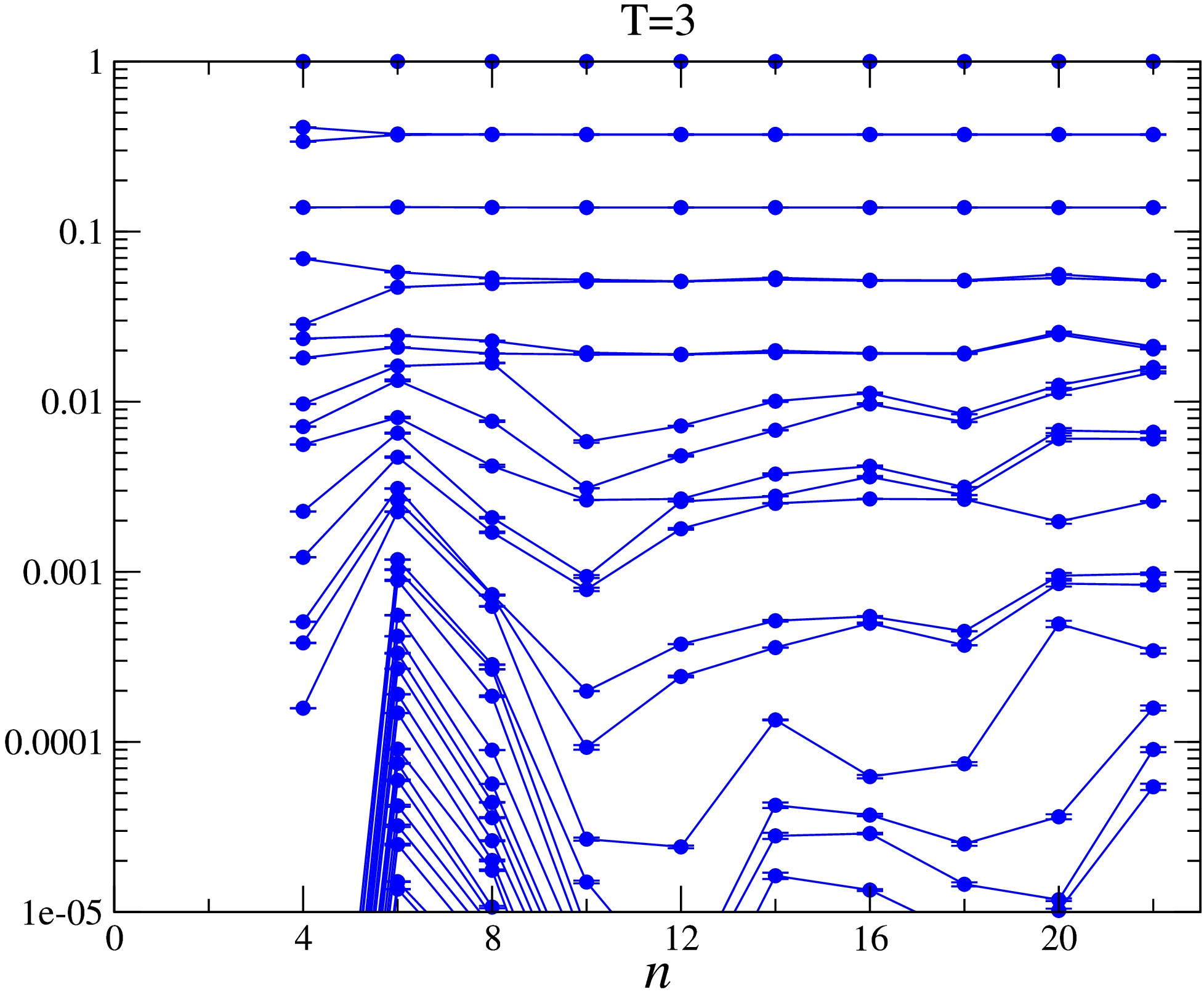} 
\end{center}
\caption{\label{fig:SVspectra} 
The upper panels show RG flow of the singular value spectra for the ensemble averaged tensors.
In the lower panel we show RG flow of 
the ensemble average of the singular values obtained from each sample
of the renormalized tensor. 
The 50 largest singular values normalized by the largest one are plotted 
as a function of the RG steps for three different temperatures 
of $T=2, T_{\rm c}$ and $3$ with $D_{\rm{svd}}=7$, $N_r=1$. 
We use $500$ statistics for each.
}
\end{figure}

Next we investigate the RG flow of the renormalized tensors. 
Since the position-dependent noise method can provide exact results even with the finite bond dimension, 
it is interesting to see if the renormalized tensor 
follows a physically correct RG flow in large volume limit. 
In the upper panels of Fig.~\ref{fig:SVspectra} we show the RG flow of the singular value spectra 
of the ensemble averaged renormalized tensor 
\begin{align}\label{eq:Tave}
\frac{1}{N} \sum_{\ell=1}^N T^{(n)[\ell]}.
\end{align}
We use three different temperatures below and above $T_{\rm c}$.  
At $T > T_{\rm c}$, 
the renormalized tensor becomes to be dominated by only one singular mode, 
and the other singular values are strongly 
suppressed. 
On the other hand at $T < T_{\rm c}$, we find that 
two degenerate singular values dominate the renormalized tensors. 
At $T=T_{\rm c}$, all the singular values are densely distributed. 
These behaviors are consistent with the expectations 
from the real space RG transformation of the Ising model. 
A similar observation has already been made 
in other improved tensor network algorithms  
that can remove short distance correlations 
and take into account the environment effects of the tensor networks~\cite{evenbly2015tensor, Hauru:2017jbf}. 
It should be noted here that 
we use exactly the same coarse-graining process as the original TRG. 
Nevertheless our method seems to reproduce a physically correct RG flow of the renormalized tensors 
despite its simple and easy algorithm. 
One of the reasons for this is that in our method all the singular modes are included in the renormalized tensors hence
the long-range correlation are manifestly taken into account.
As seen in Eq.~\eqref{eq:Tave}, it is also important to take the ensemble average for the renormalized tensor first then one should see its singular values.
To make this point clear,
we investigate the RG flow of the singular values from an individual sample of renormalized tensor,
and then it turns out to obey the same spectral flow as in the TRG
as shown in lower panels of Fig.~\ref{fig:SVspectra}.  
Therefore we see that the ensemble averaging procedure reduces the statistical noise and makes the renormalized tensor to be close to the correct RG flow.

\subsection{Common noise method}
\noindent 
In this subsection we basically show results of the common noise method on normal lattice. 
We shall explicitly mention whether the normal lattice or the checkerboard lattice when necessary.

Even for the common noise method, it is natural to use
Eq.~\eqref{eq:fV} to calculate the free energy density.
The average of the partition function $Z_V$, however, has
the systematic errors due to the noise cross contamination effect
as discussed in Sec.~\ref{sec:coarse_graining_common_noise}.
Furthermore, as explained in Sec.~\ref{sec:position_dependent_noise},
the partition function itself has an exponentially broad distribution which suffers from the long tail effect 
and the error may not be correctly estimated. 
Therefore it is not necessary to adhere $f_V$ in Eq.~\eqref{eq:fV} and
one may pursue a different way to compute the free energy density. 
One of possibility, we propose here, is to calculate an ensemble average of 
the log of the partition function, 
\begin{align}\label{eq:fbar}
\bar{f}_V = -\frac{T}{V} \langle \log{Z^{(n)}} \rangle, 
\quad 
\langle \log{Z^{(n)}} \rangle = \frac{1}{N} \sum_{\ell=1}^N \log{Z(T^{(n)[\ell]})}, 
\end{align}
where the statistical error is estimated from the variance. 
Needless to say, $\bar{f}_V$ also contains the systematic error and $\bar{f}_V\neq f_V$.
However, it turns out that
$\bar{f}_V$ has a much better control of the systematic error than $f_V$ as we see later 
[see paragraph including Eqs.~\eqref{eq:A}-\eqref{eq:limit_f_V^ex}].
Again, it should be noted here that there may exist some negative samples of $Z(T^{(n)[\ell]}) < 0$ in the common noise method as well. 
Since 
individual samples are required to calculate $\bar{f}_V$ in Eq.~\eqref{eq:fbar}, 
we have to exclude negative samples from the measurement, 
and hence there exists an additional systematic uncertainty in the common noise method. 
As shown later the appearance probability of negative samples depends on 
both the physical system and the noise parameters.  
In order to clarify the presence of such an additional systematic uncertainty 
and study its impact we consider the following distinction when we present the results.
If we do not observe any negative $Z$ samples in the ensemble, 
then we indicate the result with a filled symbol.    
On the other hand, 
if there are any negative $Z$ samples in the ensemble, 
we exclude these samples from the analysis and the corresponding result is indicated by an empty symbol.
In this case, the mean value has the additional systematic uncertainty. 

\begin{figure}[tbph]
\begin{center}
\includegraphics[clip,width=0.5\textwidth]{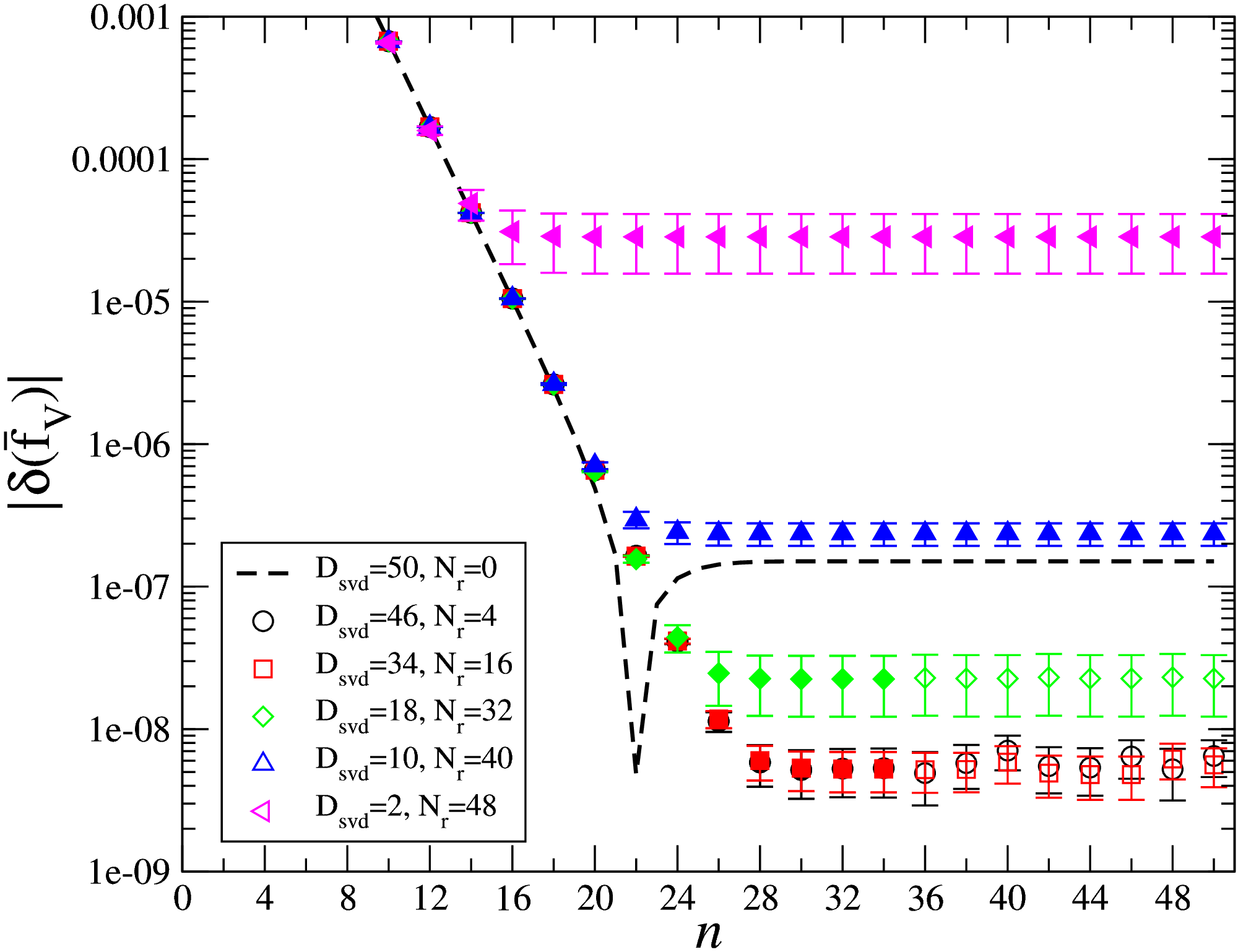} 
\end{center}
\caption{\label{fig:D50} 
The coarse-graining step dependence of relative deviation between the common noise method and the exact result 
for the free energy density $\displaystyle |\delta (\bar{f})|=|(\bar{f}-f^{\rm exact})/f^{\rm exact}|$ at the critical temperature $T=T_{\rm c}$.  
The noise dimension is varied while keeping the total bond dimension $D_{\rm cut}=D_{\rm svd}+N_r=50$ fixed. 
The number of statistics is $N=50$ for all cases.
The TRG result ($D_{\rm svd}=50$, $N_r=0$) is also shown for comparison.
The data with filled symbols indicate that we do not observe any negative sample. 
The data with empty symbols indicate that there are some negative samples in the ensemble
that are excluded from the analysis.  
}
\end{figure}

First we study the volume dependence of the relative deviation for $\bar{f}_V$ at the critical temperature $T=T_{\rm c}$ for several parameters of noise vectors 
as shown in Fig.~\ref{fig:D50}, 
where the total number of bond dimensions $D_{\rm{cut}}= D_{\rm{svd}}+N_r$ is fixed to 50. 
We use 50 statistics for each parameter. 
In comparison, the results obtained from the original TRG ($D_{\rm svd}=50, N_r=0$) are also shown. 
We observe a plateau (a fixed-point tensor) at $n \gtrsim 28$ for all the parameters 
as in the case of the original TRG. 
On the contrary to the position-dependent noise method, 
this plateau indicates that there exists a systematic error that can not be reduced by increasing the statistics.
As discussed in Eq.~\eqref{eq:contact}, 
this is the noise cross contamination effect due to a multiple use of the same noise vectors. 
The detailed analysis of this residual systematic error $\propto1/N_r$ will be discussed near the end of this subsection.
As for the negative samples, 
since this is caused by a large fluctuation due to the random noise,  
the appearance probability of negative samples can be suppressed by increasing $N_r$ as shown in Table~\ref{tab:rate}. 
In practice, we find that with fixed $N_r =\mathcal{O}(1)$ 
a negative sample [$Z(T^{(n)[\ell]})<0$] seems to be generated after reaching a plateau. 
In order to avoid such the additional systematic uncertainty
$D_{\rm cut}$ should be increased in accordance with increasing $n$. 
We also find that in contrast to the position-dependent noise method, 
the number of the negative samples does not drastically change as increasing $n$. 

As for the accuracy of the common noise method, 
we obtain a significantly improved accuracy for smaller values of $N_r=4$ and 16. 
On the other hand, the results with larger values of $N_r$ (with smaller $D_{\rm svd}$) have worse accuracy. 
This is because the tensor approximation via the SVD is not good for such smaller $D_{\rm{svd}}$ as shown in Eq.~\eqref{eq:Mdecompose},   
which should also lead to a significant increase of the noise contamination error as decreasing $D_{\rm svd}$.
Thus in order to minimize the error one should consider an optimization for the parameters of $D_{\rm svd}$ and $N_r$.
By changing these two parameters, we could minimize the noise contamination error. 
We find that typically a few number of $N_r \ll D_{\rm svd}$
is sufficient to obtain better accuracy than the original TRG. 

\begin{table}[ht]
\begin{tabular}{r|crrrrrrrrrr}
$n$ & $1$ - $30$ & 32 & 34 & 36 & 38 & 40 & 42 & 44 & 46 & 48 & 50 \\\hline 
$D_{\rm svd}=46, N_r=\phantom{0}4 $ & 0 & 1 & 2 & 2 & 3 & 6 & 3 & 3 & 4 & 6 & 5 \\
$D_{\rm svd}=34, N_r=16$ & 0 & 0 & 0 & 3 & 3 & 4 & 2 & 2 & 2 & 4 & 2 \\
$D_{\rm svd}=18, N_r=32$ & 0 & 0 & 0 & 1 & 1 & 1 & 2 & 1 & 1 & 2 & 1 
\end{tabular}
\caption{Number of the negative samples at $D_{\rm cut}=D_{\rm svd}+N_r=50$ for total $N=50$ samples.
}\label{tab:rate}
\end{table}

\begin{figure}[tbph]
\begin{center}
\includegraphics[clip,width=0.5\textwidth]{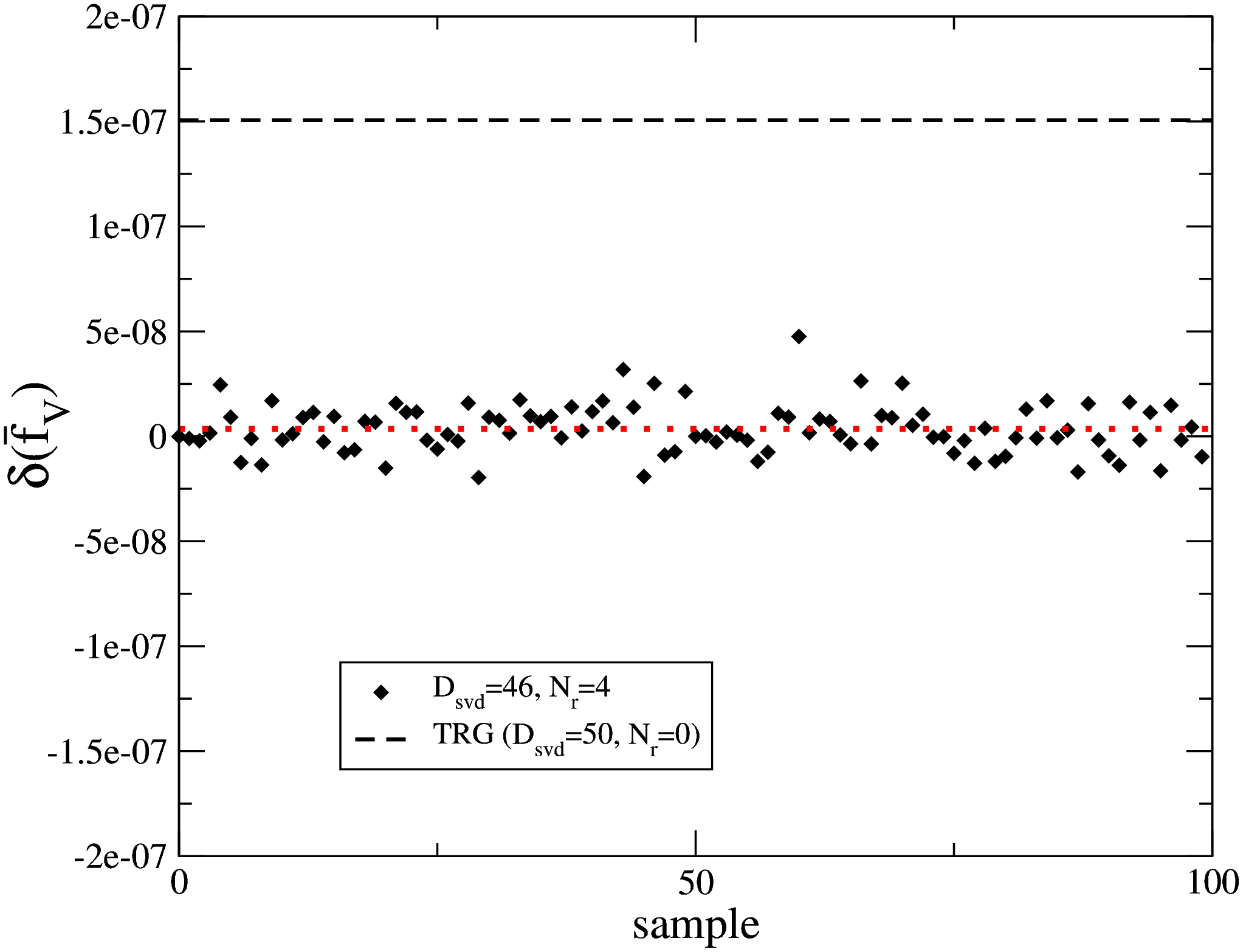} 
\end{center}
\caption{\label{fig:scatter} 
Sample dependence of the relative errors for the free energy density $\delta(\bar{f}_V)$ at the step $n=30$ and $T=T_{\rm c}$.
The red dotted and black dashed lines represent the mean value of $N=100$ samples 
and the absolute value of TRG result, respectively.
}
\end{figure}

Figure \ref{fig:scatter} shows a scatter plot for 100 samples of the results with $D_{\rm{svd}}=46$ and $N_r=4$. 
We find that all the samples have much better accuracy than the original TRG 
and the fluctuation is well controlled. 
In fact the plot indicates that the systematic error of TRG with $D=50$, of $\mathcal{O}(\Lambda_{51}$),
is larger than the statistical fluctuation of our method with $D_{\rm svd}=46$ and $N_r=4$, of $\mathcal{O}(\Lambda_{47}/\sqrt{4}$),
which is estimated from Eq.~\eqref{eq:M}.
Since our noise method utilizes all SVD modes in the tensor decomposition
and the noises are spatially correlated, such a small fluctuation is not surprising,
as long as the additional systematic error due to the noise correlation is under control.
The mean value is stable against the changes of the statistics 
compared to the position-dependent noise method, 
that is, only a few statistics is sufficient to obtain a reliable estimate. 

\begin{figure}[tbph]
\begin{center}
\includegraphics[clip,width=0.5\textwidth]{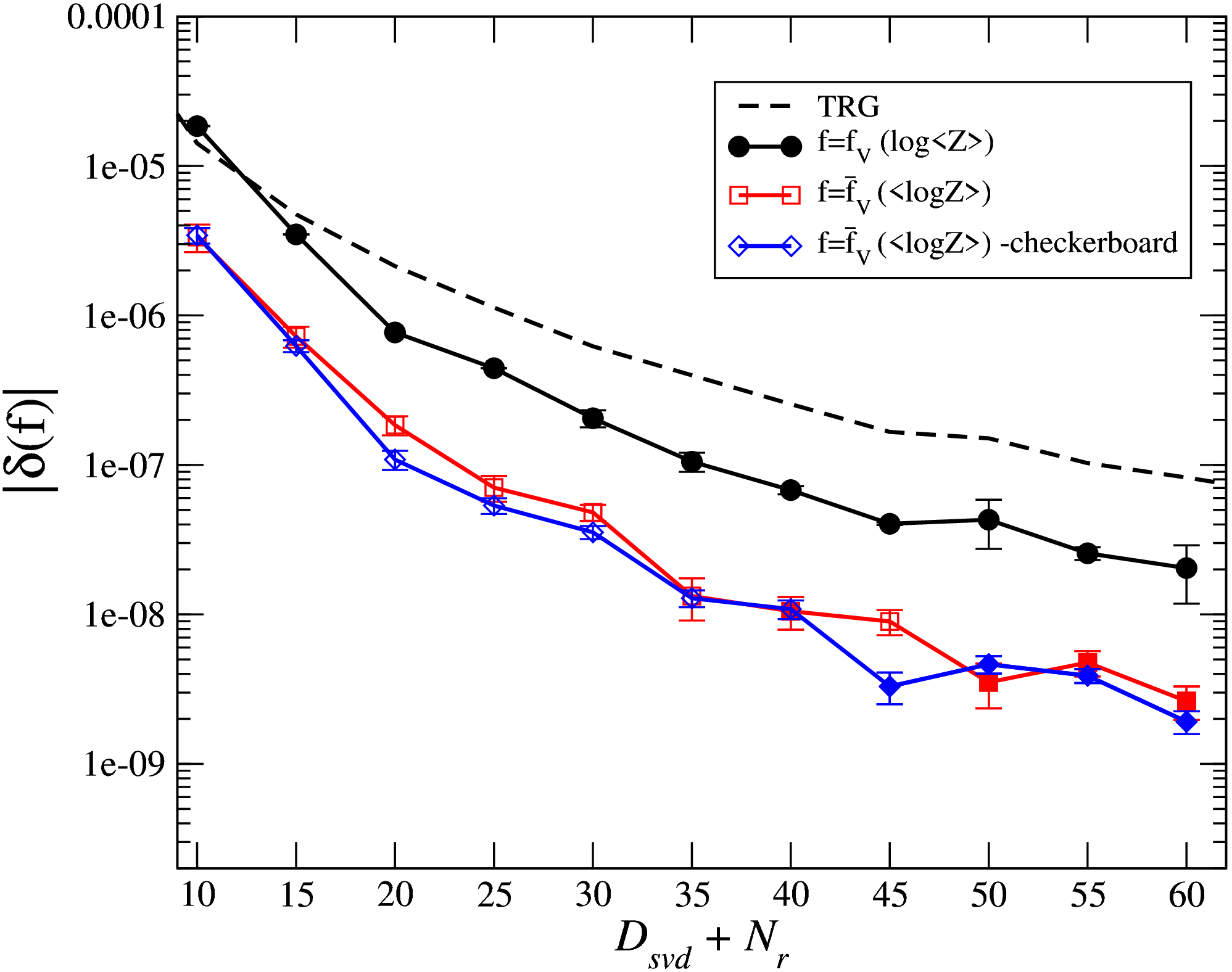} 
\end{center}
\caption{\label{fig:Ddep_S30} 
$D_{\rm cut}$ dependence of the relative errors of the free energy density on 
$V=2^n$ with $n=30$ and $N_r=4$ at $T=T_{\rm c}$ for the common noise method with normal lattice (black circle and red square) and checkerboard lattice (blue diamond). 
We use $N=100$ samples for each parameter.
For comparison the TRG results with the bond dimension $D_{\rm cut} = D_{\rm svd} + N_r $ ($N_r=0$) are also plotted.  
}
\end{figure}

Figure \ref{fig:Ddep_S30} shows 
the $D_{\rm cut}$ dependence of the relative error of the free energy at $T=T_{\rm c}$ with fixed $N_r=4$ at $n=30$. 
Similar to the original TRG, 
we observe a monotonically decreasing of $|\delta(\bar{f})|$ as increasing the bond dimension $D_{\rm cut}$. 
We compare two methods of normal and checkerboard lattices in the figure. 
We observe a slightly better accuracy for the checkerboard lattice
as expected. 
We also compare 
two evaluations of the free energy, $f_V$ in Eq.~\eqref{eq:fV} and $\bar{f}_V$ in Eq.~\eqref{eq:fbar}.
As shown in the figure, a significant improvement in $\bar{f}_V$ rather than $f_V$ is observed, 
which means that the systematic error of $\bar{f}_V$ is much smaller 
than that of $f_V$.\footnote{Note that the statistical error for $f_V$ 
may not be correctly estimated for such small statistics due to the exponentially broad distribution.}

We shall briefly discuss a possible reason for the error reduction mechanism in $\bar{f}_V$. 
Let us consider a spin-0 toy model instead of the original Ising model, 
namely a renormalized tensor $A$ has a bond dimension $D_{\rm cut}=1$.
For simplicity we assume that $A$ consists of two terms as
\begin{align}
A = A_{\rm ex}(1 + \delta_A), 
\quad \quad 
\label{eq:A}
\end{align}
where the first term $A_{\rm ex}$ is an exact part, 
and the second term $\delta_A$ represents a statistical fluctuation due to random noises with $\langle \delta_A \rangle =0$ where
$\langle \cdots \rangle$ denotes the ensemble average as given in \eqref{eq:fbar}. 
Assuming $|\delta_A|\ll 1$, the partition function $Z^{(n)}$ is then given as 
\begin{align}
Z^{(n)} \equiv A^V = A_{\rm ex}^V
\left( 1 + V \delta_A + \frac{V(V-1)}{2} (\delta_A)^2 \right)
+ \mathcal{O}((\delta_A)^3), 
\end{align}
and its mean value is 
\begin{align}
\langle Z^{(n)} \rangle =& A_{\rm ex}^V
\left(1 + \frac{V(V-1)}{2} \langle \delta_A^2 \rangle \right)
+ \mathcal{O}((\delta_A)^3).
\end{align}
In the case of $f_V$ 
it is given as the log of the mean value 
\begin{align}
f_V =& - \frac{T}{V} \log{\left(\langle Z^{(n)}\rangle\right)}
\simeq  
f_V^{\rm{ex}} 
-T \frac{(V-1)}{2} \langle \delta_A^2\rangle, 
\label{eqn:fV_fex}
\end{align}
where the exact free energy is given by $f_V^{\rm{ex}} =-T \log{A_{\rm ex}}$. 
On the other hand, in the case of $\bar{f}_V$ 
it is given as 
\begin{align} 
\bar{f}_V =& - \frac{T}{V} \langle \log{Z^{(n)}} \rangle 
\simeq 
f_V^{\rm{ex}} +
T \frac{\langle \delta_A^2 \rangle}{2}.
\label{eqn:barfV_fex}
\end{align}
Equations (\ref{eqn:fV_fex}) and (\ref{eqn:barfV_fex}) tell us that $f_{V}$ and $\bar{f}_V$ are not simply equal to
the exact one due to the noise cross contamination effect at leading order $\langle\delta_A^2\rangle$, however,
according to the argument in Eq.~(\ref{eq:contact})
the systematic error scales with $\langle\delta_A^2\rangle\propto 1/N_r$, therefore
such a systematic error can be removed for sufficiently large $N_r$,
\begin{align} 
\lim_{N_r\to\infty} f_V
=&
\lim_{N_r\to\infty} \bar{f}_V 
=f_V^{\rm ex}.
\label{eq:limit_f_V^ex}
\end{align} 
For finite $N_r$, 
there is the systematic error $\langle \delta_A^2 \rangle$ for both methods, but
the coefficient for $\bar{f}_V$ is much smaller than that of $f_V$,
which is proportional to volume\footnote{In the case of the position-dependent noise method, 
the noise cross contamination effect $\langle \delta_A^2 \rangle$ should be 
replaced with $\langle \delta_{A^i} \rangle \langle \delta_{A^j} \rangle$ $(i \neq j)$, 
where $\delta_{A^i}$ is a noise fluctuation on site $i$. 
A broad distribution with volume dependence discussed in the previous section is also understood from this analysis.}
[see Eq.~(\ref{eqn:fV_fex})].
In fact, the error reduction in $\bar{f}_V$ is observed in all the parameter regions 
for both the normal and checkerboard lattices. 
As for the additional systematic uncertainty in $\bar{f}_V$ due to negative samples, 
it can be avoided by taking sufficiently large $N_{r}$, 
so Eq.~\eqref{eq:limit_f_V^ex} should hold.

\begin{figure}[tbph]
\begin{center}
\includegraphics[clip,width=0.5\textwidth]{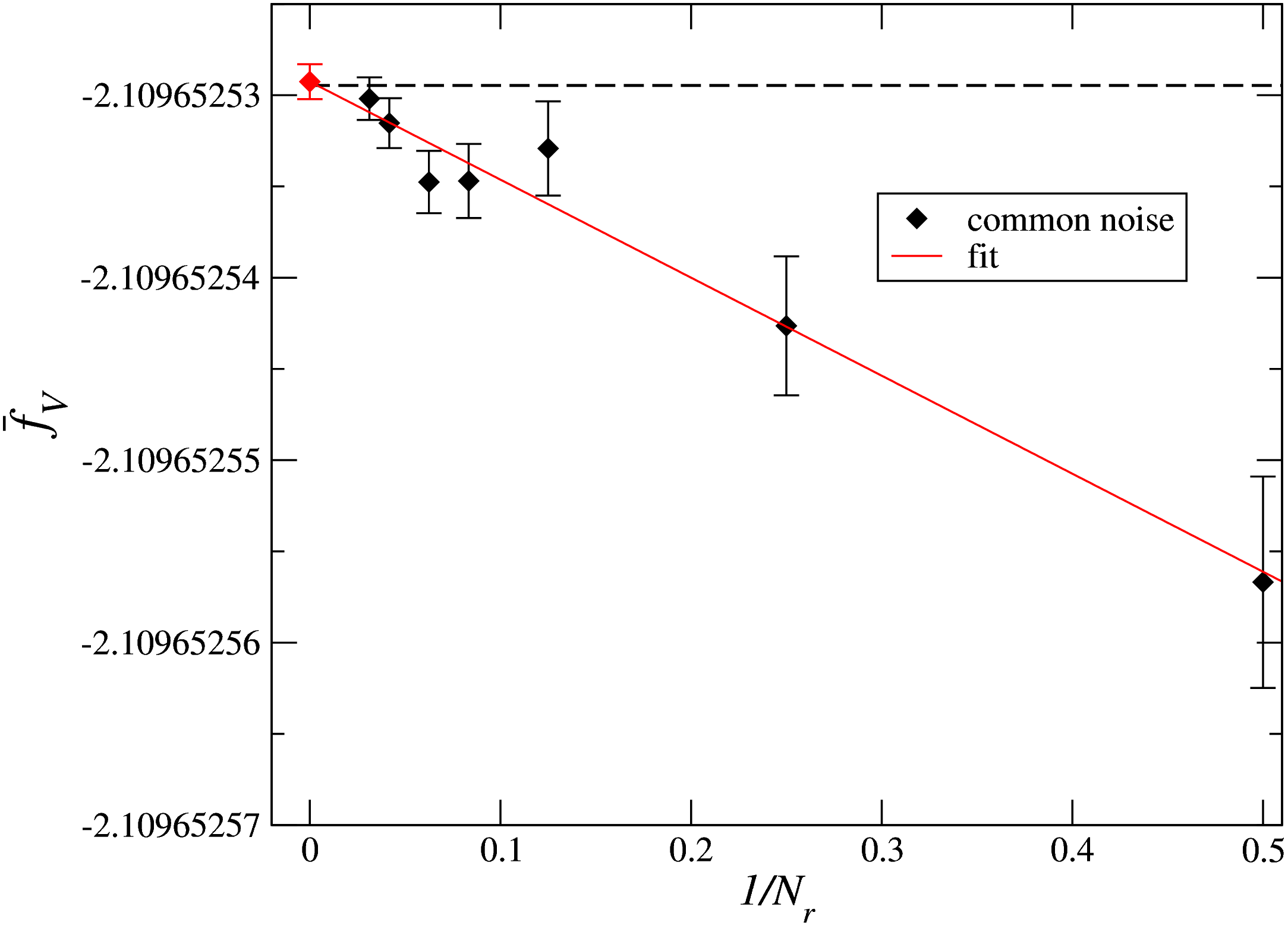} 
\end{center}
\caption{\label{fig:D20_S20} 
$N_r$ dependence of the free energy density $\bar f_V$
with fixed $D_{\rm svd}=20$ at the coarse-graining step $n=20$, $T=T_{\rm c}$, and $N=5000$ statistics for each parameters. 
The dashed line represents the analytic result on $V=2^{20}$.
}
\end{figure}

\begin{figure}[tbph]
\begin{center}
\includegraphics[clip,width=0.5\textwidth]{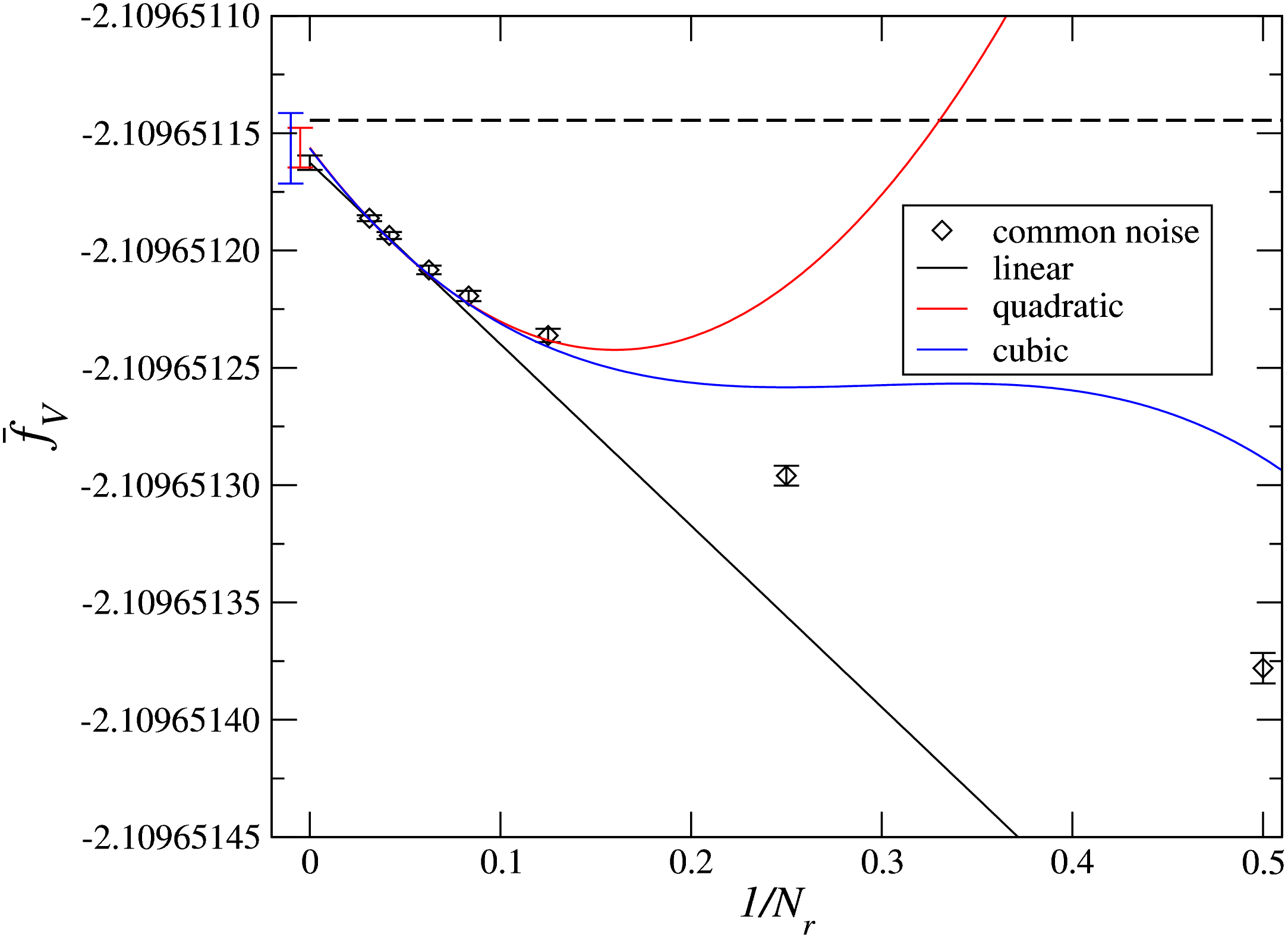} 
\end{center}
\caption{\label{fig:D20_S40} 
$N_r$ dependence of the free energy density  $\bar f_V$
with fixed $D_{\rm svd}=20$ at the coarse graining step $n=40$, $T=T_{\rm c}$, and $N=5000$ statistics for each parameters. 
The dashed line represents the analytic result on $V=2^{40}$.
}
\end{figure}

Next we study the scaling property of the systematic error. 
As shown in the previous section, 
the residual systematic error in the common noise method 
should be given by the noise cross contamination effect in Eq.~\eqref{eq:contact},
and it is important to study the $N_r$ dependence of the results. 
Figure \ref{fig:D20_S20} shows the free energy density at the coarse-graining step $n=20$ as a function of $1/N_r$
with fixed $D_{\rm svd}=20$. 
$N_r$ is varied in the range $2-32$.
As shown in the figure, since we observe a clear linear dependence of $1/N_r$,  
we can carry out a fit analysis to obtain a result in the $N_r \to \infty$ limit. 
We consider a linear fit function of $\bar{f}_V = d_0 + d_1/N_r$ 
and use all the data for a fit.
The fit result is given as $d_0 = -2.1096525293(10)$ with $\chi^2/\rm{d.o.f.}=0.7$, 
which is consistent with the analytic value at $V=2^{20}$.
For larger volumes 
we show the results at the step $n=40$ with fixed $D_{\rm svd}=20$ 
as a function of $1/N_r$ in Fig.~\ref{fig:D20_S40}. 
In contract to the results with $n=20$, 
we see a curvature so that higher order corrections of $1/N_r$ are required to fit
these data. 
We consider higher order polynomial fit functions 
\begin{align} \label{eq:poly}
\bar{f}_V = d_0 + d_1 \frac{1}{N_r} + d_2 \frac{1}{N_r^2} + d_3 \frac{1}{N_r^3}.
\end{align}
The fit results are also shown in the figure and Table~\ref{tab:D20_S40}.  
Our results are well fitted by polynomial functions in $1/N_r$. 
We obtain $\bar{f}_{2^{40}} = -2.1096511637(28) $ for cubic functions in $N_r=\infty$ limit, 
which is consistent with the analytic result at $V=2^{40}$.
Thus we obtain much better results than the original TRG,  
and the noise cross contamination effect is found to be well controlled 
by the $1/N_r$ corrections.\footnote{
We note that in the case of $n=40$
we observe negative samples in the entire region of $N_r$, 
and hence the results have the additional systematic uncertainty. 
Thus there is a possibility that the curvature in $n = 40$ 
is caused by the systematic effect due to the negative $Z$ samples.
To correctly study a $1/N_r$ scaling 
we would need more data for much larger $N_r$ region where there is no negative $Z$ sample.}

We note here that since this contamination is the only systematic error, 
the $1/N_r$ scaling is a universal feature of the common noise method, 
which should not depend on the details of the systems and the coarse-graining processes. 
Therefore this scaling property is quite different from the original TRG. 
As for the MCTN, 
since the common noise method has not been examined in~\cite{ferris2015unbiased, 2017arXiv171003757H},  
it is not obvious if one could see a clear scaling property of the systematic error.  
It may be interesting to study the scaling property in the common noise method for the MCTN as well.

\begin{table}[ht]
\begin{tabular}{c|cccccc}
& $d_0$ & $d_1 \times 10^{-7}$ & $d_2 \times 10^{-6}$ & $d_3 \times 10^{-6}$ & $\chi^2/$d.o.f. & 
fit range
\\\hline
linear & $-2.1096511625(31)$ & $-7.74(72)$ & - & - & $6 \times 10^{-3}$ & $N_r \geq 16$ \\
quadratic & $-2.1096511562(85) $ & $-1.08(34) $ & 3.4(3.1) & - & $0.3$ & $N_r \geq 12$ \\
cubic & $-2.109651156(15)\phantom{0}$ & $-1.08(7.6) $ & 4(11) & -4(50) & $0.3$ & $N_r \geq 8$ \\
\end{tabular}
\caption{Fit results for $\bar f_V$ at $n=40$
using a polynomial fit form in Eq.~\eqref{eq:poly}.
The value of $\chi^2/$d.o.f. and the fit ranges are also tabulated.
}\label{tab:D20_S40}
\end{table}

\begin{figure}[tbph]
\begin{center}
\includegraphics[clip,width=0.28\textheight]{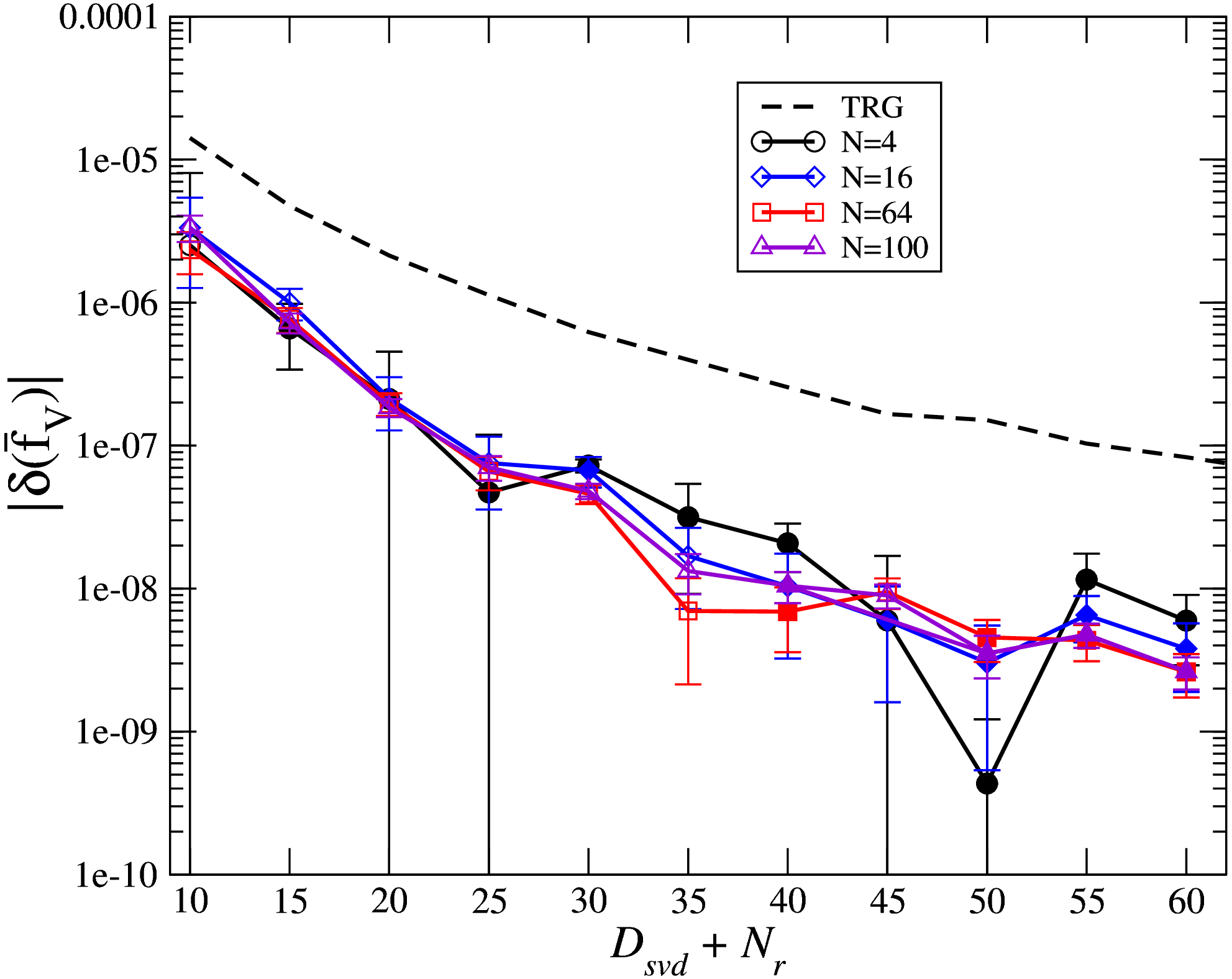} 
\quad \quad 
\includegraphics[clip,width=0.285\textheight]{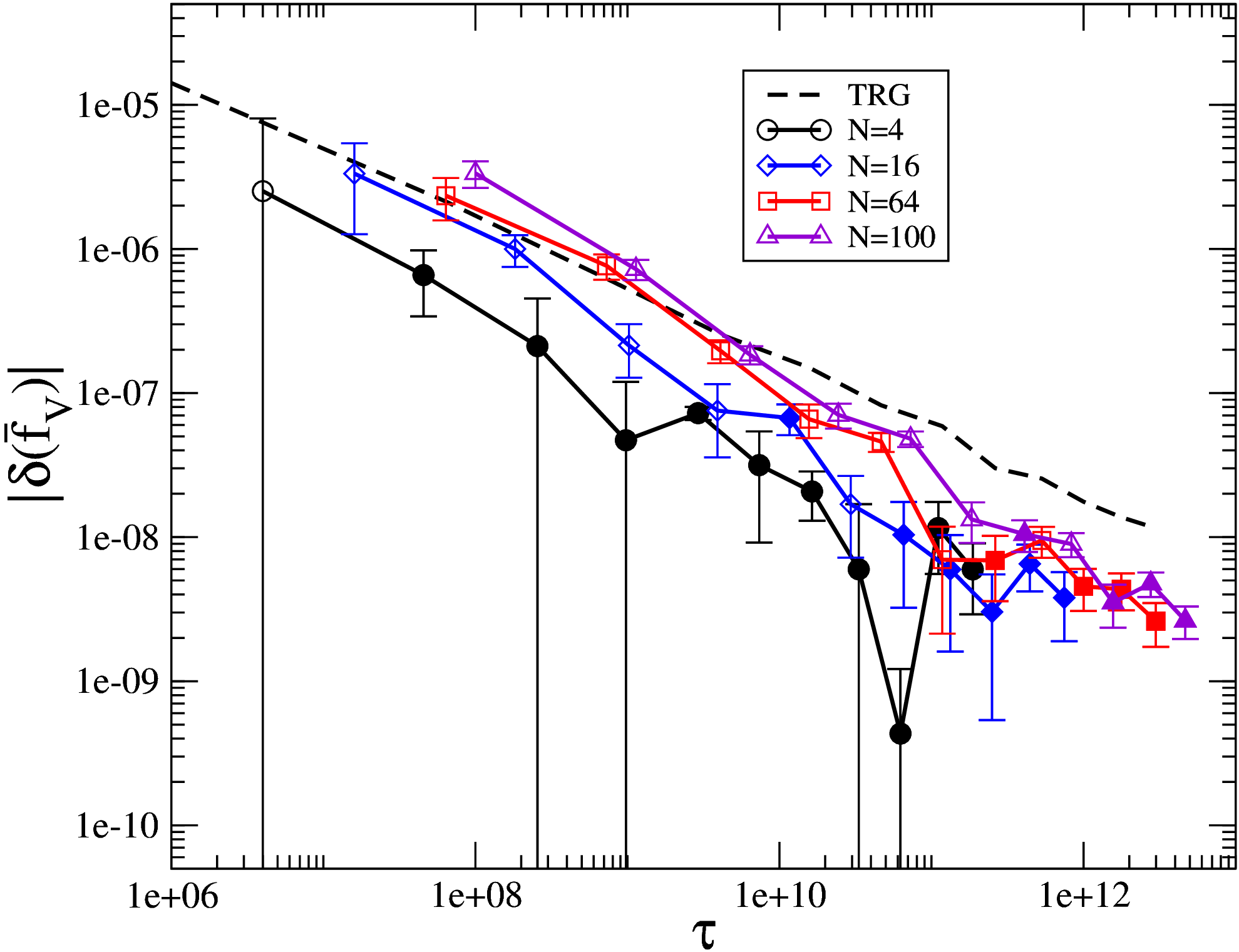} 
\end{center}
\caption{\label{fig:comparison} 
$D_{\rm cut}$(left panel) and $\tau$(right panel) dependence of the relative errors $|\delta(\bar{f}_V)|$ 
for various statistics on $V=2^{30}$ with fixed $N_r=4$ at $T=T_{\rm c}$.
}
\end{figure}

Finally we shall discuss the sample size dependence of the performance 
in comparison with the TRG. 
In order to contextualize the numerical cost of our method, which depends on the number of statistics $N$, 
we estimate a computational time $\tau$ as \cite{Adachi:2019paf}
\begin{align}
\tau = 
\begin{cases}
D_{\rm cut}^6  & \ {\rm for \ TRG} \\
N D_{\rm cut}^6 & \ {\rm for \ common \ noise \ method.}
\end{cases}
\end{align}
The left panel of Fig.~\ref{fig:comparison} shows the 
relative error as a function of $D_{\rm cut}$ for various $N$.  
The mean values for different $N$ are consistent with each other.  
A substantial reduction of the systematic error is found to be statistically significant 
even for a small $N$. 
In the right panel of Fig.~\ref{fig:comparison}, 
we show the same results as a function of $\tau$.
For smaller values of $\tau < 10^{10}$
the numerical cost becomes comparable with the TRG  with increasing $N$. 
On the other hand, for larger $\tau$ the common noise method 
has better performance even for large values of $N$.
We note that the actual elapsed time 
can be even more reduced by utilizing parallel computing, 
since individual samples can be independently generated.

We note that other thermodynamical quantities are also calculated based on the stochastic approach. 
As an example, we show the results for the specific heat that can be obtained 
by numerical differences of $\bar{f}_V$ in Appendix. 
It should be also noted that the common noise method can be straightforwardly implemented 
in other tensor networks without any substantial increase in computational cost
and any iterative process. 
Thus it is interesting to apply this method to other models. 
In the next section, we test this method with an advanced algorithm to explore a possibility of further improvement.

\section{\label{sec:sGilt-TNR}Application to other coarse-graining algorithms}
As shown in the previous section,
our methods provide a very simple and 
useful way 
to reduce and control the systematic errors. 
This is a distinguished feature of our methods 
in sharp contrast to other improved tensor network algorithms that are designed to 
exhibit a correct RG flow. 
In addition, our methods can be easily applied to 
other tensor networks 
as long as 
the truncated SVD
is used in a matrix decomposition.
In this regard let us demonstrate how our methods can work in other tensor network algorithms. 
Here as an example, we consider Gilt-TNR~\cite{Hauru:2017jbf}.
It is known that the Gilt procedure also provides a simple algorithm to systematically 
truncate the bond dimensions of tensor networks.
Furthermore Gilt-TNR, which is a combination of Gilt and TRG, 
also exhibits a physically correct RG flow of the renormalized tensors with better precisions. 
From the practical point of view, 
it is interesting to see whether there still is room for numerical improvement by combining with our noise methods. 

\begin{figure}[tbph]
\begin{center}
\includegraphics[clip,width=0.8\textwidth]{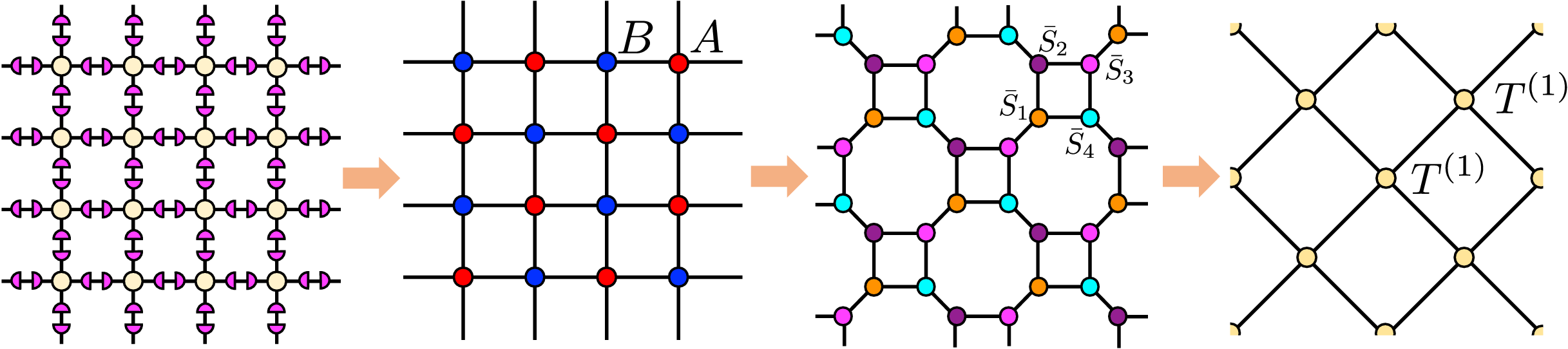} 
\end{center}
\caption{\label{fig:sGilt-TNR} 
Coarse graining
of Gilt-TNR in combining with the common noise method. 
The original tensor $T$ resides on every site, and 
four rank-reduced matrices obtained via Gilt are inserted into four links on a plaquette.
These matrices are decomposed by full SVDs and absorbed into
neighboring tensors $T$, by which 
two rank-reduced tensors of $A$ and $B$ are obtained.
These $A$ and $B$ are decomposed by the common noise methods.  
In the initial lattice four different noise vectors ${\bm \eta}_{1,2,3,4}$ are 
distributed to the four sites.
The tensor contraction for one plaquette is carried out, 
and a renormalized tensor $T^{(1)}$ is obtained.
}
\end{figure}

We explain an implementation of the common noise method to Gilt-TNR as follows. 
In the case of 2D Ising model on square lattice, 
a tensor network on a plaquette is approximated by inserting four matrices $R_{1,2,3,4}$ 
to four links of the plaquette via Gilt (see Fig.~\ref{fig:sGilt-TNR}). 
Then the original tensor $T$ on each site is modified to $A$ or $B$ by absorbing the neighboring matrices $R_{1,2,3,4}$ via SVDs,
by which a part of short distance correlations is removed
and the bond dimensions $D_{\rm cut}$ can be reduced to $D'_{\rm cut}$ 
which depends on the threshold parameter $\epsilon$ in Gilt as well as the dynamics.  
Here we simply apply the common noise method when decomposing two matrices $A$ and $B$.
Since the tensor network consisting of $A$ and $B$ has a checkerboard structure,
we can employ the common noise method on checkerboard lattice as explained in the previous section.
Moreover, after coarse graining the tensor network still retains a homogeneity on a checkerboard lattice
as seen in Fig.~\ref{fig:sGilt-TNR}. 
For the numerical test we use the original source code for 2D Ising model on square lattice \cite{Gilt}
with small modifications to suit our purposes.

\begin{figure}[tbph]
\begin{center}
\includegraphics[clip,width=0.5\textwidth]{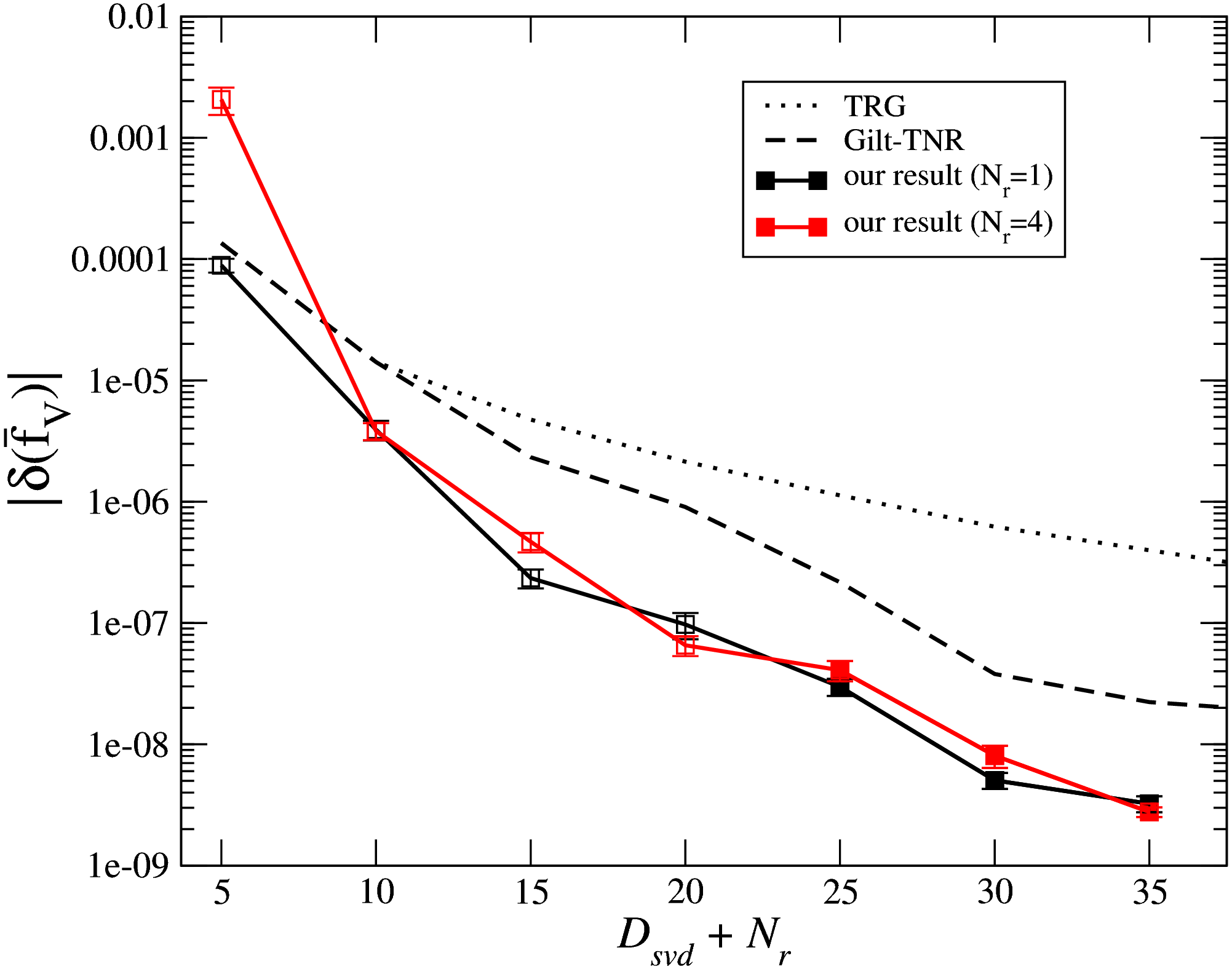} 
\end{center}
\caption{\label{fig:sGilt-TNR_S50} 
Relative deviation of the free energy at $T=T_{\rm c}$ 
with fixed values of $N_r=1$ (black), $N_r=4$ (red), 
and the threshold parameter $\epsilon=8\times 10^{-7}$.  
The number of statistics is $100$ for each result. In comparison, 
the results for TRG(dotted) and Gilt-TNR(dashed) are also shown with same parameters.  
}
\end{figure}

We present a benchmark result 
in Fig.~\ref{fig:sGilt-TNR_S50} 
which shows the relative errors $|\delta(\bar{f}_V)|$ 
as a function of $D_{\rm cut}$ 
at $T=T_{\rm c}$ with $N=100$ on $V=2^{51}$ lattice.  
For a comparison we also show the results for Gilt-TNR 
as well as the original TRG.  
We commonly choose 
the Gilt threshold parameter $\epsilon=8\times 10^{-7}$ as given in \cite{Hauru:2017jbf}. 
As shown in the figure, 
our noise methods systematically improve the accuracy. 
Our results with even smaller bond dimensions can achieve a relative error of $\mathcal{O}(10^{-9})$, 
 that is a typical precision limit of Gilt-TNR for this range of parameters. 
We note that the computational cost per sample is the same order of Gilt-TNR, 
while the reduced number of the bond dimension via Gilt depends 
not only on the threshold parameter $\epsilon$ but also on samples due to the random noises.  
Thus our method is shown to be effective even for an improved algorithm without any changes of 
the original tensor networks.

\section{\label{sec:conclusion}Conclusion}

\noindent
Following the idea of the MCTN,
we have proposed a new stochastic method by utilizing random noise vectors combining with the singular value decomposition, 
where the rank-reduced tensor manifestly contains all the singular modes thanks to the random noise vectors.
In the method we generate tensor ensembles, 
and the partition functions and any related physical quantities are statistically calculated. 
We have tested two types of the noise distribution for 2D Ising model. 

In the case of the position-dependent noise method 
since there is no systematic error, our result is exact in the sense that 
no matter how accurate result one could obtain by increasing the statistics while keeping the bond dimension finite. 
We note that the computational cost is as expensive as $\mathcal{O}(ND_{\rm cut}^6 V)$, 
which should be comparable to the cost of TRG-related methods for a system without translation invariance.
We also confirm that the RG flow of the singular value spectra are consistent 
with the expectation from the real space RG transformation despite its simple and easy algorithm.
On the other hand, in the case of the common noise method, the computational cost scales as $\mathcal{O}(ND_{\rm cut}^6\log V)$ 
and
we obtain a better accuracy than the original TRG with a limited number of statistics. 
While there exists a residual systematic error due to a multiple use of the noise vectors, 
this error is found to be under control by a model independent $1/N_r$ scaling, 
where $N_r$ is the number of the noise dimensions. 
Thus our stochastic method actually improves the error evaluation method as well as the numerical accuracy.
This model independent property is in sharp contrast to other tensor network algorithms using the truncated SVD. 
It should also be emphasized that our method is very simple and does not 
require any iterative process, 
so it is easily implemented even to other improved tensor networks. 
As a nontrivial example, we have applied our method to the Gilt-TNR,
where an even more error reduction has been obtained.
An interesting future direction will be an application to more 
complicated system with higher dimensionality. 

\begin{acknowledgments}
H. O. is supported in part by JSPS KAKENHI Grants No.~21K03554 and No.~22H00138.
S. T. is supported in part by JSPS KAKENHI Grants No.~20H00148, No.~21K03531, and No.~22H01222.
M. T. is supported by U.S. DOE Grants No.~DE-SC0010339, and No.~DE-SC0021147.
This work was supported by MEXT KAKENHI Grant-in-Aid for Transformative Research Areas A “Extreme Universe” No. 22H05251

\end{acknowledgments}

\section*{Appendix: Specific heat}\label{app:c}
\begin{figure}[tbph]
\begin{center}
\includegraphics[clip, height=0.2\textheight]{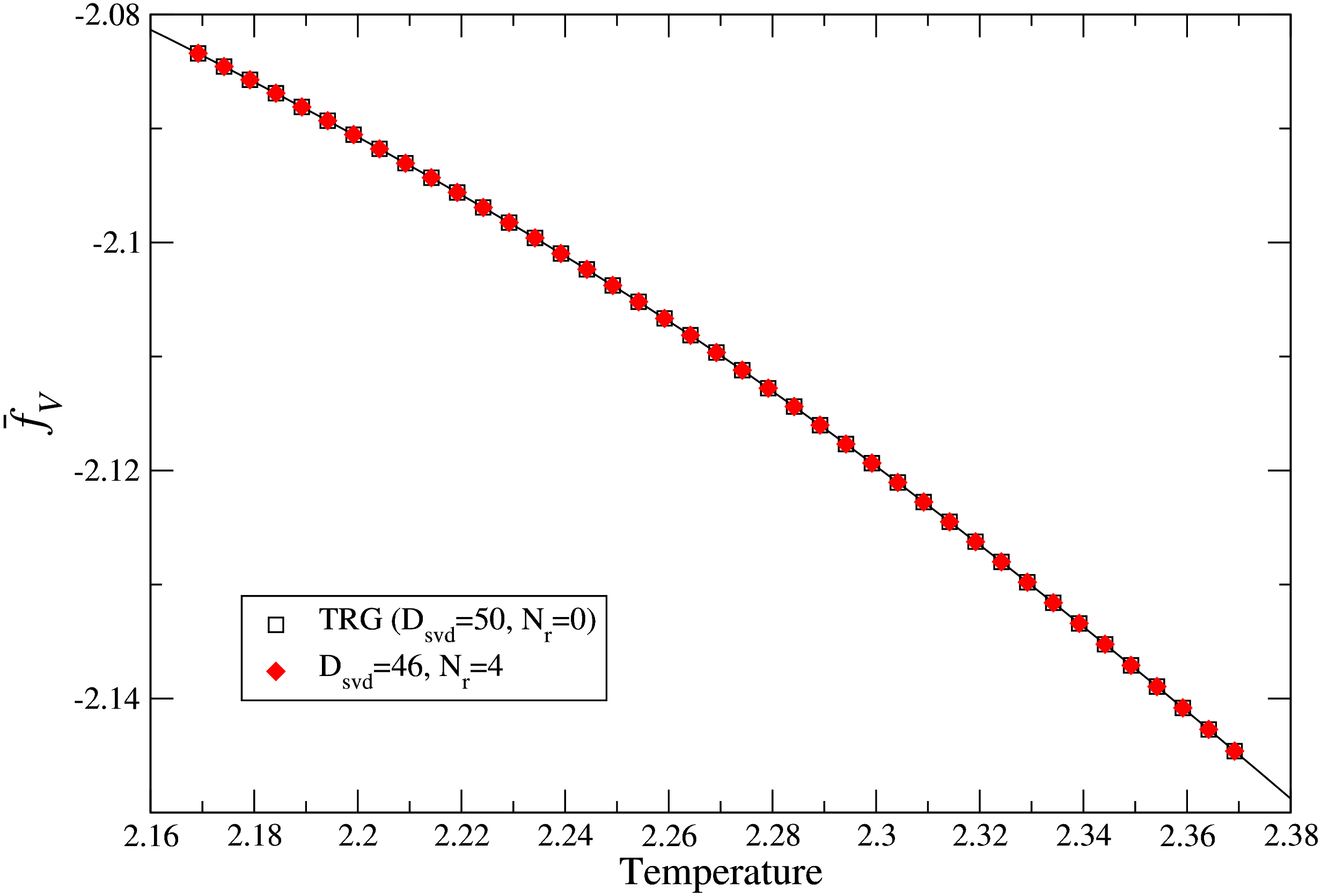} 
\quad \quad 
\includegraphics[clip,height=0.2\textheight]{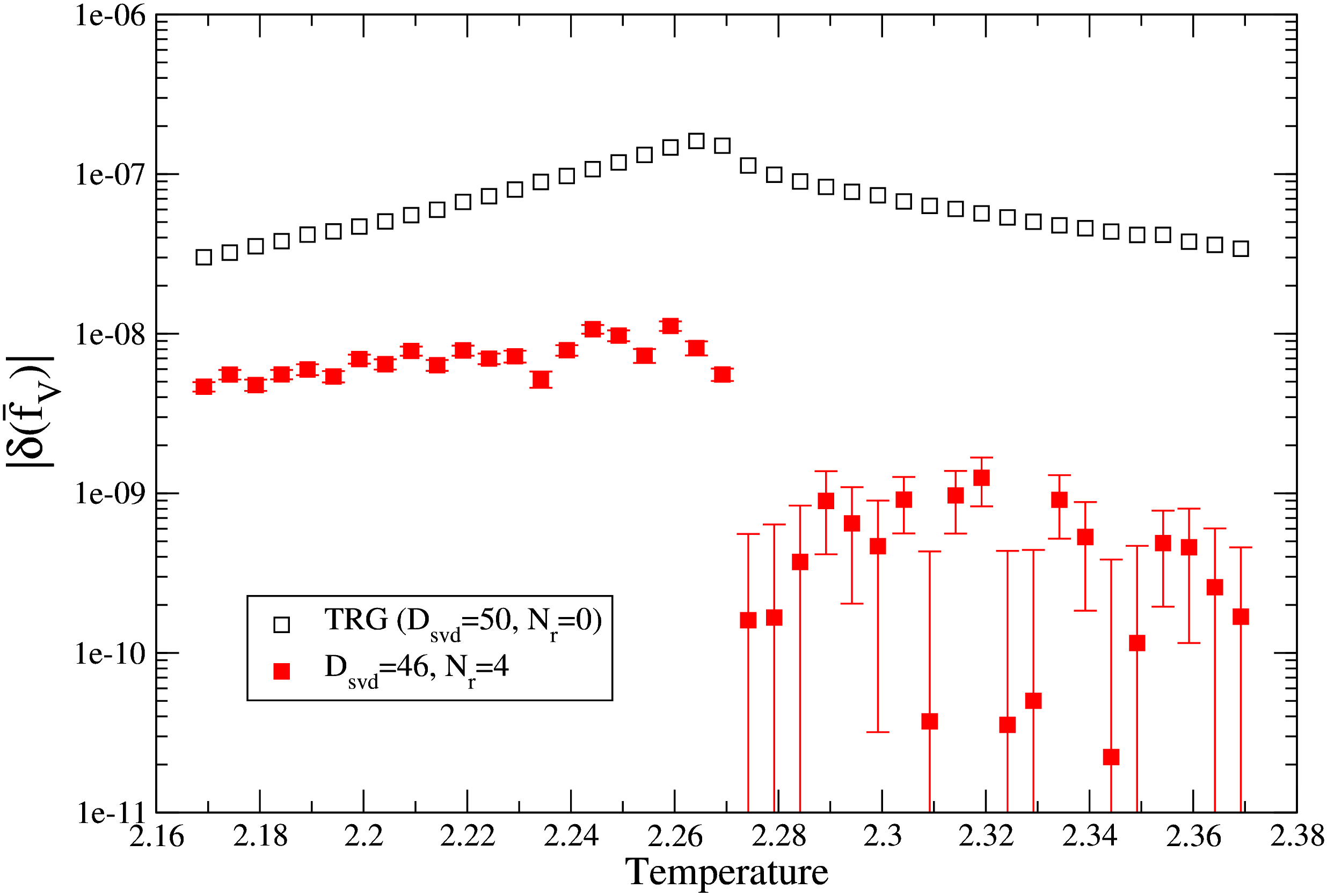} 
\end{center}
\caption{\label{fig:f} 
Temperature dependence of $\bar{f}_V$(left panel) and $|\delta(\bar{f}_V)|$(right panel) 
on $V=2^{30}$ with $N=100$ statistics 
is shown in comparison with the results for the TRG (square). 
The solid curve in the left panel represents the Onsager's exact results.
}
\end{figure}

As an example of thermodynamical quantities, we study the specific heat  based on the stochastic approach for the common noise method. 
Figure \ref{fig:f} shows the temperature dependence of the free energy $\bar{f}_V$ and the absolute value of the relative error $|\delta(\bar{f}_V)|$ 
around the critical temperature $T=T_{\rm c}$. 
From this results, we calculate the specific heat $C_V$ from numerical differentials of $\bar{f}_V$ 
with a finite interval $\Delta T$ of temperature $T$. \\

\begin{figure}[tbph]
\begin{center}
\includegraphics[clip, height=0.2\textheight]{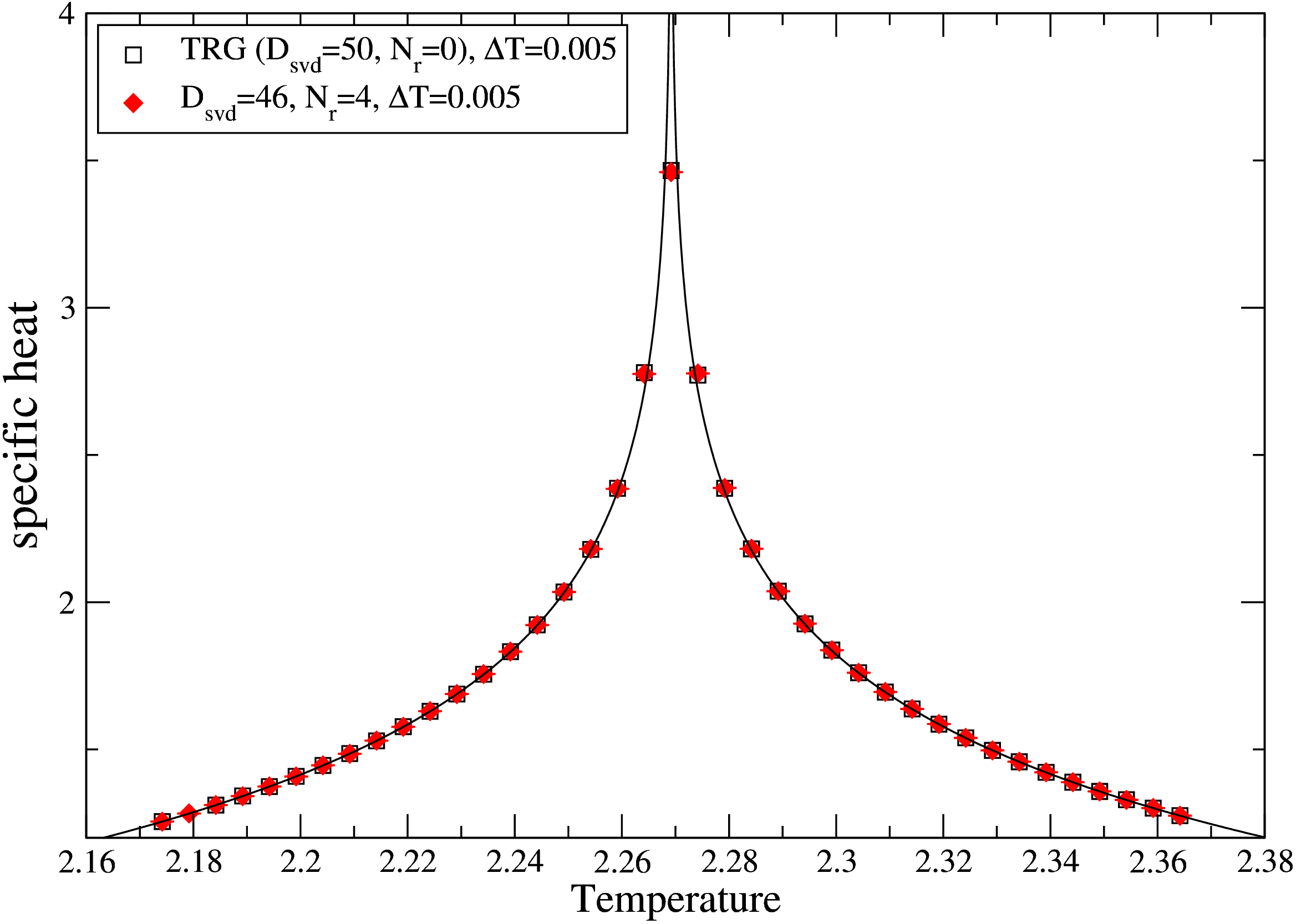} 
\quad \quad 
\includegraphics[clip,height=0.2\textheight]{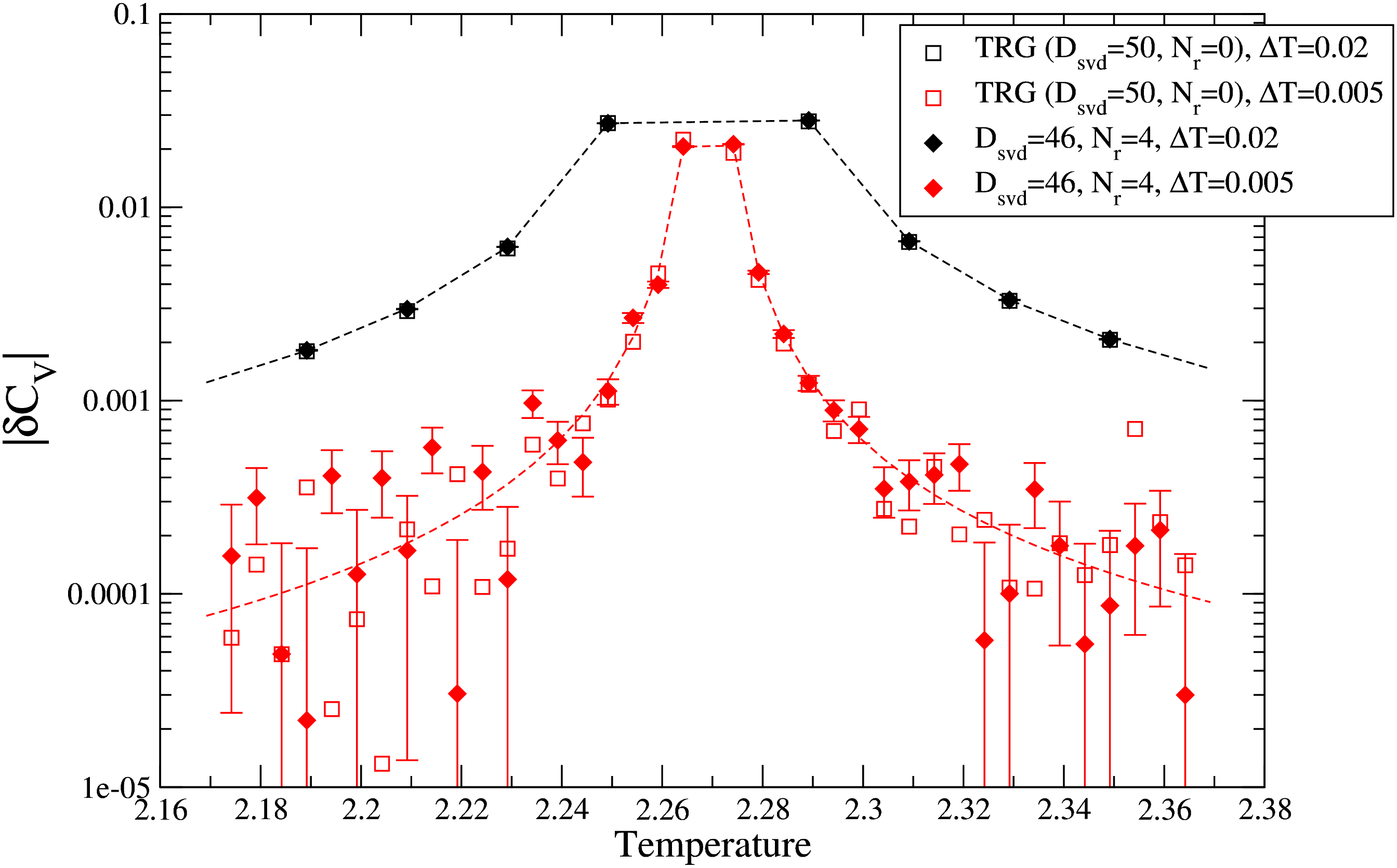} 
\end{center}
\caption{\label{fig:c} 
Temperature dependence of $C_V$(left panel) and $|\delta C_V|$(right panel) 
on $V=2^{30}$ with $N=100$ statistics is shown in comparison with the results for the TRG (square). 
The solid curve in the left panel represents the Onsager's exact results.
The dashed curves in the right panel represent
the analytic results on the finite volume obtained from numerical differences of the 
free energy $f_V^{\rm analytic}$ with a finite interval $\Delta T$.
}
\end{figure}

In Fig.~\ref{fig:c} we show $T$ dependence of $C_V$ and the absolute value of the relative error $|\delta C_V|$ 
around the critical temperature $T=T_{\rm c}$, 
where $\delta C_V = (C_V - C^{\rm exact})/C^{\rm exact}$ 
is a relative error from the Onsager's exact result for the specific heat $C^{\rm exact}$. 
We show the results for $|\delta C_V|$ obtained with two values of $\Delta T$, 0.02 and 0.005, 
which indicate the significance of the finite-$\Delta T$ effect on the results from both TRG and the stochastic methods. 
Decreasing $\Delta T$ can in principle reduce this effect but in stead increases the significance of other errors, the roundoff effects on both results 
and the statistical error on the result from the stochastic method, as seen for $\Delta T = 0.005$.
The enhancement of the statistical error on numerical differential is known. 
In fact, it has been observed in \cite{Nakamura:2018enp} that when using a stochastic method, 
the precision of the free energy is better while the specific heat obtained by the numerical difference is comparable with that of the TRG.
The increase of $|\delta C_V|$ near $T_{\rm c}$ should be because of the finite-volume effect.
Since these dominant finite-volume and finite-$\Delta T$ effects are the same for both TRG and the stochastic method, the accuracy of these methods are comparable.

\bibliography{sTRG}

\end{document}